\documentclass[12pt,amsmath,amssymb,bibliography]{revtex4}

\usepackage{amsfonts}
\usepackage{amsmath}
\usepackage{amssymb}
\usepackage{array}
\usepackage{bm}
\usepackage{braket}
\usepackage{breqn}
\usepackage{cleveref}
\usepackage{color}
\usepackage{dcolumn}
\usepackage{epsfig}
\usepackage{epsf}
\usepackage{epstopdf}
\usepackage{extarrows}
\usepackage{float}
\usepackage{graphicx}
\usepackage{graphics}
\usepackage{harpoon}
\usepackage{indentfirst}
\usepackage{mathrsfs}
\usepackage{multirow}
\usepackage{pifont}
\usepackage{xcolor}

\newcommand {\sla}[1]{ #1 \!\!\!/}

\newcommand{\RM}[1]{\textrm{\uppercase\expandafter{\romannumeral#1}}}

\renewcommand{\thefootnote}{\fnsymbol{footnote}}

\begin{document}

\title{$\gamma W$-exchange contributions in neutron $\beta$ decay in the forward-angle limit}

\author{
Hui-Yun Cao$^{1}$\protect\footnotemark[1]\protect\footnotetext[1]{E-mail: caohy@hbnu.edu.cn},
Hai-Qing Zhou$^{2}$\protect\footnotemark[2]\protect\footnotetext[2]{E-mail: zhouhq@seu.edu.cn} \\
$^1$ School of Physics and Electronic Science, Hubei Normal University, HuangShi 435002, China\\
$^2$ School of Physics, Southeast University, NanJing 211189, China}

\date{\today}

\begin{abstract}
In this work, the contributions from $\gamma W$-exchange in neutron $\beta$ decay are estimated at the amplitude level. Using a general form for the electromagnetic (EM) form factors (FFs) of the proton, the EM FFs of the neutron, and the weak FFs of the $Wnp$ interaction as inputs, we present analytical expressions for the inner part of the $\gamma W$-exchange amplitude under the forward angle limit. The differences and relations between our method and those used in the references are discussed. To compare our numerical results with those provided in the references, we consider three types of FFs as examples. The numerical results show that when the favored EM FFs are used, our result for the contribution from the Fermi part is consistent with those reported in the references, while our results for the Gamow-Teller parts are about 7\% and 13\% larger than those reported in Ref.~\cite{Hayen-2021}.
\end{abstract}

\maketitle

\section{Introduction}

Neutron $\beta$ decay is triggered by the electroweak interaction, where a free neutron decays into a proton, accompanied by the emission of an electron and an electron antineutrino ($n \rightarrow p e \bar{\nu}_{e}$). Due to its clean background, neutron $\beta$ decay serves as a powerful laboratory for extracting the Cabibbo-Kobayashi-Maskawa (CKM) matrix $V_{ud}$ in the Standard Model (SM) \cite{PDG-2024}, which is important for checking the universality of the quark mixing \cite{CKM-Unitary}. Additionally, it can be used to extract the structure of the proton, such as the axial-vector coupling constant $g_A$, which provides necessary input for nuclear physics, particle physics, astrophysics, and cosmology \cite{Role-of-gA-1,Role-of-gA-2,Role-of-gA-3}.

For neutron $\beta$ decay, the value of $V_{ud}$ is extracted from the experimental data via the following theoretical relation
\begin{align}
|V_{ud}|^2 \propto  \frac{1}{\tau_n(1+3\lambda^2)}\frac{1}{1+\text{RCs}},
\end{align}
where $\tau_n$ is the neutron lifetime, $\lambda\equiv g_A/g_V$ with $g_{A,V}$ being the axial-vector and vector coupling constants of the $Wnp$ interaction, and RCs denote the electroweak radiative corrections, which are related to the structures of the proton and neutron and should be calculated theoretically.

Protons and neutrons are bound by the strong interaction, and their structures are described by Quantum Chromodynamics (QCD). Due to the non-perturbative nature of QCD, it is still difficult to calculate RCs from first principles. For the RCs in free neutron $\beta$ decay, as summarized in the literature \cite{Seng-2021}, the best starting point is Sirlin's representation. Sirlin \cite{Sirlin-1967} decomposes the RCs into outer (known as model-independent) and inner (known as model-dependent) parts as follows:
\begin{align}
\text{RCs} = \frac{\alpha_e}{2\pi} \bar{g}(E_m) + \Delta_R^V,
\end{align}
where $\alpha_e$ is the fine structure constant, and the outer correction $\bar{g}(E_m)$ is a universal function that represents the extreme infrared part of the radiative correction and is exactly calculable. The inner correction $\Delta_R^V$ is related to short-range and hadronic structure effects, which are the main sources of uncertainty in extracting $V_{ud}$ and $\lambda$ from experimental data. The analyses in Refs. \cite{Sirlin-1978, Sirlin-2004, Sirlin-2006} give the inner RCs as
\begin{align}
\Delta_R^V &= \frac{\alpha_e}{2\pi} \Big[ 3\ln \frac{m_Z}{m_{\rho}} + \ln\frac{m_Z}{m_W} + \tilde a_g \Big] + \delta_{\text{HD}}^{\text{QED}} + 2~\square_{\gamma W}^V,
\end{align}
where the first two terms on the right-hand side are the resummation of leading quantum electrodynamic (QED) logarithms, $\tilde a_g$ is the perturbative QCD (pQCD) correction, and $\delta_{\text{HD}}^{\text{QED}}$ represents contributions from the leading logs of higher-order QED effects. The term $\square_{\gamma W}^V$ denotes the RCs originating from the $\gamma W$-exchange diagrams, which can be expressed as
\begin{align}
\square_{\gamma W}^V &= \frac{\alpha_e}{2\pi} \Big[ \frac{1}{2} \ln \frac{m_W}{m_A} + C_{\text{Born}} + \frac{1}{2} \text{Ag} \Big],
\end{align}
where the first term in the square brackets represents the large electroweak logarithm, with $m_A$ as an effective IR cutoff scale, $C_{\text{Born}}$ is the Born elastic contribution to the low-energy part of the $\gamma W$-exchange diagram, and Ag is the pQCD correction.

Refs. \cite{Towner-1992, Hayen-2021} calculated $C_{\text{Born}}$ in the forward angle limit (FAL), where the proton and neutron are treated as isospin doublet states, and special combinations of their electromagnetic (EM) form factors (FFs) are used as inputs to estimate the contributions. In this study, we discuss the inner RCs of $\gamma W$-exchange contributions to the amplitude of neutron $\beta$ decay, and then relate these corrections to $C_{\text{Born}}$. In our calculation, we take the EM FFs of the proton and neutron as independent inputs since they are measured independently in different experiments. Furthermore, we use a very general form for these FFs as input to provide the analytic expressions for the $\gamma W$-exchange amplitude. These analytical expressions can be used not only to estimate the corrections to the lifetime of the neutron, but also to estimate the corrections to all polarized measurements. Additionally, they can be utilized when the measurements of the EM FFs of the proton and neutron are improved in the future.

The paper is organized as follows. In Sec. II, we provide some basic formulas, including the expressions for the amplitudes, the approximations used in Refs. \cite{Towner-1992, Hayen-2021}, the existing problems, and the parameterization of the input FFs. In Sec. III, we express the amplitude as a combination of 16 terms in the form of Pauli spinors, and then present the analytical expressions for the coefficients of these 16 terms in the FAL. The relations between these coefficients and $C_{\text{Born}}$ are also discussed. In Sec. IV, we take three types of FFs as inputs to present the numerical comparisons between our results and those reported in the literature. A detailed discussion of these numerical results and the conclusion are provided.

\section{$\gamma W$-exchange contributions}
\subsection{The One-$W$-exchange and $\gamma W$-exchange amplitudes}
\begin{figure}[htbp]
\centering
\includegraphics[height=7cm]{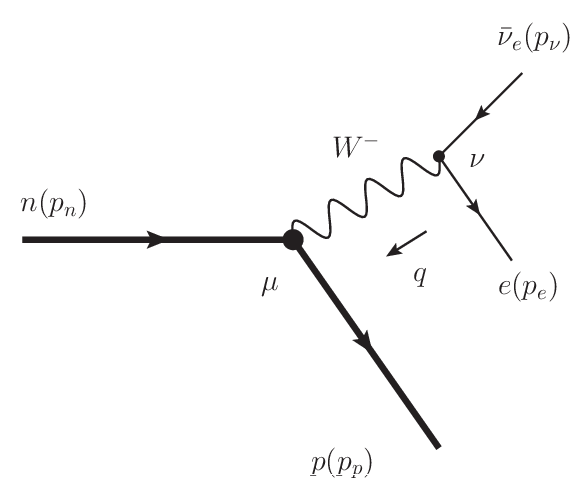}
\caption{Diagram for $n\rightarrow p e \bar{\nu}_e$ with one-W-exchange.}
\label{figure:one-W-exchange-diagram}
\end{figure}

\begin{figure}[htbp]
\centering
\includegraphics[height=7cm]{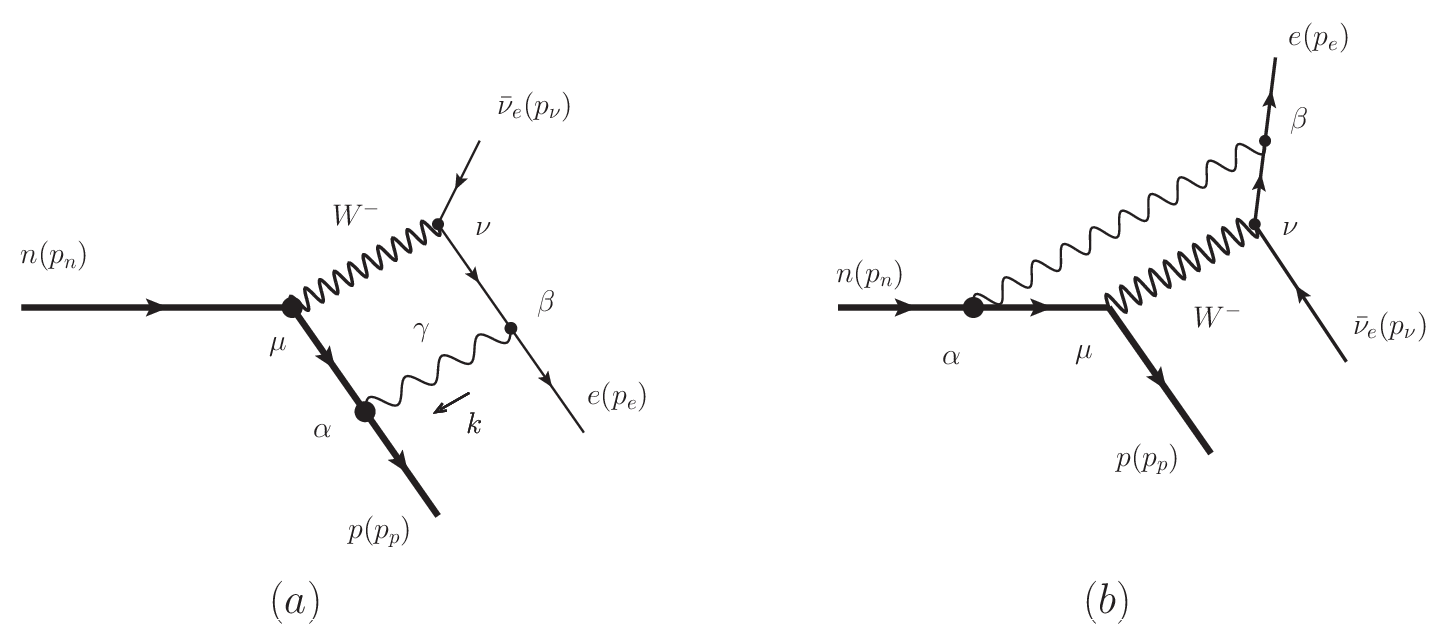}
\caption{Diagrams for $n\rightarrow p e \bar{\nu}_e$ under $\gamma W$-exchange, where only the elastic intermediate
states are considered.}
\label{figure:gamma-W-exchange-diagram}
\end{figure}

At the hadronic level, the tree diagram for the neutron $\beta$ decay can be described by Fig.~\ref{figure:one-W-exchange-diagram}, where only one-$W$-exchange diagram contributes. When considering the RCs from $\gamma W$-exchange with the elastic intermediate state, the diagrams shown in Fig.~\ref{figure:gamma-W-exchange-diagram} contribute, and the corresponding amplitudes can be expressed as follows:
\begin{align}
\mathcal{M}^{W} &= -i\Big[\bar{u}(p_e,m_e)\Gamma^{\omega}_{W \nu e} u(p_{\nu}, m_{\nu})\Big] ~\Big[\bar{u}(p_p,m_p) \Gamma^{\mu}_{Wnp}(q) u(p_n,m_n)\Big]\frac{-ig_{\mu\omega}}{q^2-m_W^2},\notag\\
\mathcal{M}^{\gamma W}_{(a)} &= -i\int\frac{d^4 k}{(2\pi)^4}\Big[ \bar{u}(p_e,m_e) \Gamma^{\beta}_{\gamma ee}S_F(p_e+k,m_e)\Gamma^{\omega}_{W \nu e}  u(p_{\nu},m_{\nu})\Big] \notag\\
& \times\Big[\bar{u}(p_p,m_p) \Gamma^{\alpha}_{\gamma pp}(k)S_F(p_p-k,m_p) \Gamma^{\mu}_{Wnp}(q-k) u(p_n,m_n)\Big]  \frac{-ig_{\mu\omega}}{q^2-m_W^2}\frac{-ig_{\alpha\beta}}{k^2+i\epsilon},  \notag\\
\mathcal{M}^{\gamma W}_{(b)}&= -i\int\frac{d^4 k}{(2\pi)^4}\Big[\bar{u}(p_e,m_e) \Gamma^{\beta}_{\gamma ee} S_F(p_e+k,m_e)\Gamma^{\omega}_{W \nu e} u(p_{\nu},m_{\nu}) \Big] \notag\\
&\times \Big[\bar{u}(p_p,m_p)\Gamma^{\mu}_{Wnp}(q-k)S_F(p_n+k,m_n) \Gamma^{\alpha}_{\gamma nn}(k) u(p_n,m_n)\Big]  \frac{-ig_{\mu\omega}}{q^2-m_W^2}\frac{-ig_{\alpha\beta}}{k^2+i\epsilon},
\label{equation:OBE-and-TBE-amplitude}
\end{align}
where $\bar{u}(p_e,m_e), u(p_{\nu},m_{\nu}),\bar{u}(p_p,m_p)$ and $ u(p_n,m_n)$ are the spinors of the electron, antineutrino, proton and neutron with the corresponding momentum and mass, respectively. Here $q\equiv p_p-p_n$ and $k$ is momentum of photon. Additionally, the $S_F$ is given by
\begin{align}
S_F(l,m) &= \frac{i(\sla{l}+m)}{l^2-m^2+i\epsilon},
\end{align}
and the vertices are
\begin{align}
\Gamma^{\mu}_{\gamma ee} =& -ie\gamma^{\mu},~~~~~~~~\Gamma^{\mu}_{W \nu e} =i\frac{g}{2\sqrt{2}}\gamma^{\mu}(1-\gamma_5),\\
\Gamma^{\mu}_{\gamma pp}(l) =& ie\Big[F_{1}^{p}(l^2)\gamma^{\mu}+i\frac{F_{2}^{ p}(l^2)}{2m_p}\sigma^{\mu\nu}l_{\nu}  \Big],
\notag\\
\Gamma^{\mu}_{\gamma nn}(l) =& ie\Big[F_{1}^{ n}(l^2)\gamma^{\mu}+i\frac{F_{2}^{ n}(l^2)}{2m_n}\sigma^{\mu\nu}l_{\nu}  \Big],\\
\Gamma^{\mu}_{W n p}(l) =&  i\frac{g}{2\sqrt{2}}V_{ud}\Big[\Big(f_1(l^2)\gamma^{\mu}+i\frac{f_2(l^2)}{2m_N}\sigma^{\mu\rho}l_{\rho}+\frac{f_3(l^2)}{2m_N}l^{\mu}\Big)\\
&~~~~~~~~~+\Big( f_4(l^2)\gamma^{\mu}+i\frac{f_5(l^2)}{2m_N}\sigma^{\mu\rho}l_{\rho}+\frac{f_6(l^2)}{2m_N}l^{\mu} \Big)\gamma_5\Big],
\label{equation:vertex-gammaNN}
\end{align}
where $l$ is the momentum of the incoming photon or $W$ boson, $e=-|e|$ is the EM interaction coupling constant, $g$ is the SU(2) gauge coupling constant, $m_N\equiv\frac{m_n+m_p}{2}$, $F_{1,2}^{ p}(l^2)$ and $F_{1,2}^{n}(l^2)$ are the EM FFs of the proton and neutron, $f_1(l^2)$, $f_2(l^2)$, $f_3(l^2)$, $f_4(l^2)$, $f_5(l^2)$, and $f_6(l^2)$ account for vector, weak magnetism, scalar, axial vector, weak electricity, and induced pseudoscalar contributions, respectively.

In the above expressions, the on-shell FFs are used in the loop integration. One might question whether the nucleon could be off-shell in the loop diagrams, necessitating consideration of the off-shell effects. To address this, the dispersion relations (DRs) method or lattice calculation is employed to estimate the corresponding contributions in Refs.~\cite{Seng-1807,Seng-1812,Seng-2106,MaPengXiang-2024}. It is important to note that the imaginary parts of the box diagrams are independent of off-shell effects, and the full results obtained from the above expressions automatically satisfy certain DRs, as indicated in the $ep$ scattering case~\cite{Tomalak:2014sva,Cao:2020tsx}. This leads to a significant property: the differences between the real parts of the box diagrams calculated directly and those obtained using un-subtracted DRs are polynomial functions of momenta. These polynomials correspond to the subtracted terms in the DRs. In other words, both methods yield the same results when the same on-shell FFs are used as inputs, except for the presence of some polynomial terms. Furthermore, for finite $q^2$, it is advisable to apply DRs at the amplitude level rather than at the cross-section level, as the latter approach may overlook contributions from kinematic singularities, such as the $\gamma Z$-exchange contribution in $ep$ scattering case \cite{Guo-2023}.

In the literature, the RCs are usually separated into inner contributions and outer contributions. For the former, it is equivalent to keeping only the terms with Levi-Civita tensors in the amplitude of the leptonic part after the products of Dirac matrices are reduced. This means the following replacement:
\begin{align}
& i\bar{u}(p_e,m_e) \gamma^{\beta}S_F(p_e+k,m_e)\gamma^{\omega}(1-\gamma_5) u(p_{\nu},m_{\nu}) \notag\\
=& \frac{-(2p_e+k)^{\beta}L^{\omega}-k^{\omega}L^{\beta}+k\cdot L g^{\omega\beta}+i\epsilon^{\beta\omega\rho \sigma}L_{\rho}k_{\sigma}}{(p_e+k)^2-m_e^2}\notag\\
\rightarrow&  \frac{i\epsilon^{\beta\omega\rho \sigma}k_{\sigma}L_{\rho} }{(p_e+k)^2-m_e^2},
\end{align}
where
\begin{align}
L_{\rho} &\equiv \bar{u}(p_e,m_e)\gamma_{\rho} (1-\gamma_5)u(p_{\nu},m_{\nu}).
\end{align}
In the following discussion, we only focus on the inner contributions.

\subsection{Approximations used to estimate $\gamma W$-exchange contributions in literatures}

In Refs.~\cite{Towner-1992, Hayen-2021}, the scalar term $f_3$ and the weak electricity term $f_5$ are neglected, as these contributions correspond to the second-class currents (SCC). In this work, we also adopt this approximation as follows:
\begin{align}
f_3(l^2) &\equiv g_S\approx 0,  \notag\\
f_5(l^2) &\equiv g_T\approx 0.
\end{align}
In addition, the following FAL
\begin{align}
m_n\approx m_p \approx m_N, p_n\approx p_p \approx p,
\label{equation:FW-limit}
\end{align}
is used in Refs. \cite{Towner-1992,Hayen-2021} before loop integration. This means that
\begin{align}
\bar{u}(p_p,m_p)\Gamma_H^{\mu\alpha}(p_p,p_n,k) u(p_n,m_n)&\approx \bar{u}(p,m_N)\Gamma_H^{\mu\alpha}(p,p,k) u(p,m_N),
\end{align}
where $\Gamma_H^{\mu\alpha}(p_p,p_n,k)$ refers to hadronic part of  ${\cal M}^{\gamma W}_{(a,b)}$.

After applying the FAL and the properties of the Dirac equation to simplify the products of Dirac matrices in the hadronic part amplitude, the authors of Ref.~\cite{Hayen-2021} further used the following first-class current (FCC) approximation:
\begin{align}
\bar{u}(p)\left[\gamma_\mu(\not p-\not k+M) \gamma_\nu\right] u(p)&\xlongequal{\text{FCC}}-i \epsilon_{\mu \rho \nu \sigma} k^\rho \bar{u}(p) \gamma^\sigma\gamma^5 u(p),\notag\\
\bar{u}(p)\Big[ \sigma_{\mu\alpha}k^{\alpha}(\sla{p}-\sla{k}+m_N)\gamma_{\nu}  u(p)\Big] &\xlongequal{\text{FCC}}  0, \notag\\
\bar{u}(p)\Big[ \gamma_{\mu}(\sla{p}-\sla{k}+m_N) \sigma_{\nu\alpha}k^{\alpha}  u(p)\Big] &\xlongequal{\text{FCC}} 2m_N \epsilon_{\mu\rho\nu\sigma}k^{\rho}\bar{u}(p)\gamma^{\sigma}\gamma^5 u(p),\notag\\
\bar{u}(p)\left[\sigma_{\mu \beta} k^\beta(\not p-\not k+M) \sigma_{\nu \alpha} k^\alpha\right] u(p) &\xlongequal{\text{FCC}}-i\left(k^2-2 k \cdot p\right) \epsilon_{\mu \rho \nu \sigma} k^\rho \bar{u}(p) \gamma^\sigma \gamma^5 u(p),
\end{align}
which can be found in  Eqs. (B2a-B2d) of Ref.~\cite{Hayen-2021}.

\subsection{Non-uniqueness of FCC approximation after applying FAL}

As discussed above, in Ref.\cite{Hayen-2021}, the properties of the Dirac equation are used to simplify the products of Dirac matrices in the amplitudes after applying the FAL, and then the FCC approximation is applied. In our calculation, we find that such an approach may produce non-uniqueness in the results. Here, we take the left-hand sides of Eqs. (B2b) and (B2c) in Ref.\cite{Hayen-2021} as examples to illustrate the reason for the non-uniqueness. For the left-hand sides of Eqs. (B2b) and (B2c) in Ref.~\cite{Hayen-2021}, when applying the properties of the Dirac equation to $u(p)$, one obtains the following result, and after the FCC approximation, we have:
\begin{align}
\bar{u}(p)\Big[ \sigma_{\mu\alpha}k^{\alpha}(\sla{p}-\sla{k}+m_N)\gamma_{\nu}  u(p)\Big] &\xlongequal{\text{FCC}}  0, \notag\\
\bar{u}(p)\Big[ \gamma_{\mu}(\sla{p}-\sla{k}+m_N) \sigma_{\nu\alpha}k^{\alpha}  u(p)\Big] &\xlongequal{\text{FCC}} 2m_N \epsilon_{\mu\rho\nu\sigma}k^{\rho}\bar{u}(p)\gamma^{\sigma}\gamma^5 u(p),
\label{equation:1-FCC}
\end{align}
which are the same as Eqs. (B2b) and (B2c) in Ref.~\cite{Hayen-2021}. However, when applying the properties of the Dirac equation to $\bar{u}(p)$, one obtains the following result after the FCC approximation:
\begin{align}
\Big[\bar{u}(p) \sigma_{\mu\alpha}k^{\alpha}(\sla{p}-\sla{k}+m_N)\gamma_{\nu} \Big] u(p) &\xlongequal{\text{FCC}}  2m_N \epsilon_{\mu\rho\nu\sigma}k^{\rho}\bar{u}(p)\gamma^{\sigma}\gamma^5 u(p), \notag\\
\Big[\bar{u}(p) \gamma_{\mu}(\sla{p}-\sla{k}+m_N) \sigma_{\nu\alpha}k^{\alpha} \Big] u(p) &\xlongequal{\text{FCC}} 0.
\label{equation:2-FCC}
\end{align}
In the practical calculation, we find that these two approaches yield different results. The reason for this difference can be traced to the FAL, as the approximation $p_n\approx p_p\approx p$ used in the FAL leads to the left-hand sides of Eqs. (\ref{equation:1-FCC}, \ref{equation:2-FCC}).

To avoid such non-uniqueness, we do not use the FCC approximation after applying the FAL in the calculation. This means that we only apply the FAL to the hadronic part of the amplitude and neglect the contributions from $f_3,f_5$ when discussing the inner contributions.

\subsection{The form  of the form factors}
In practical calculation, the FFs  are introduced to simulate real physical interactions.  For the EM FFs, the isovector and isoscalar FFs are usually used \cite{Towner-1992,Hayen-2021} as  follows:
\begin{gather}
F_{1,2}^{S}\equiv F_{1,2}^{(0)} \equiv F_{1,2}^{(p)}+F_{1,2}^{(n)}, \\
F_{1,2}^{V}\equiv F_{1,2}^{(1)} \equiv F_{1,2}^{(p)}-F_{1,2}^{(n)}.
\end{gather}
When isospin symmetry is assumed, the contributions from $F_{1,2}^{V}$ are zero, and only the terms $F_{1,2}^{(S)}$ contribute. Due to this property, a dipole form  was directly introduced in Ref.~\cite{Towner-1992} to parameterize the FFs as follows:
\begin{align}
F_{1}^{(S)}(l^2)+F_{2}^{(S)}(l^2)&\equiv G_M^{(0)} =(\mu_p+\mu_n) \Big(\frac{\Lambda_\gamma^2}{l^2-\Lambda_\gamma^2}\Big)^2,
\label{equation:Towner-FFs-of-nuclear}
\end{align}
where $\mu_p$ and $\mu_n$ are the anomalous magnetic moments of the proton and neutron, respectively.

For the weak FFs after assuming $\text{SU}(2)$ symmetry and utilizing the conserved-vector-current hypothesis, one  obtains:
\begin{align}
f_{1,2}(l^2)& = F_{1,2}^{p}(l^2)-F_{1,2}^{n}(l^2).
\label{equation:f12-by-EM-FFs}
\end{align}
For $f_4$, the following simple form is used in Ref.~\cite{Towner-1992} as follows:
\begin{align}
f_4(l^2) &= f_{40} G_D(l^2,\Lambda_{W}^2),
\label{equation:dipole-FFs-of-Wnp}
\end{align}
with $f_{40}=g_A$ and
\begin{align}
G_D(l^2,\Lambda_W^2)\equiv\frac{ \Lambda_W^4}{(l^2-\Lambda_W^2)^2}.
\end{align}
The pseudoscalar FF $f_6$ is not considered, as it does not contribute to the $\gamma W$-exchange effects in our case.

Since the experimental data sets for the EM FFs of the proton and neutron are independent, we discuss their contributions separately in this work. This separation is also helpful when discussing the contributions beyond the FAL, where isospin symmetry breaking effects should be considered. Furthermore, the physical weak FFs may differ from the dipole form, as shown in Eq.~(\ref{equation:dipole-FFs-of-Wnp}). The results using a general form of FFs as inputs will be beneficial for further analysis when more experimental data sets are available in the future. In the practical calculation, we take the FFs as follows:
\begin{align}
F_{1}^{p}(l^2) &= F_{10}^{p} \sum_{j=1}^{N_1}  a_{1j}G(l^2,\Lambda_{1j}^{2},n_{1j}), \notag\\
F_{2}^{p}(l^2) &= F_{20}^{p} \sum_{j=1}^{N_2}  a_{2j}G(l^2,\Lambda_{2j}^{2},n_{2j}), \notag\\
F_{1}^{n}(l^2) &= F_{10}^{n} \sum_{j=1}^{N_3}  a_{3j}G(l^2,\Lambda_{3j}^{2},n_{3j}), \notag\\
F_{2}^{n}(l^2) &= F_{20}^{n} \sum_{j=1}^{N_4}  a_{4j}G(l^2,\Lambda_{4j}^{2},n_{4j}), \notag\\
f_{i}(l^2) &= f_{i0} \sum_{j=1}^{\bar{N}_i}  b_{ij}G(l^2,\bar{\Lambda}_{ij}^{2},\bar{n}_{ij}),
\label{equation:general-FFs-this-work}
\end{align}
where $F_{10}^{p}=1,F_{20}^{p}=\mu_p-1,F_{20}^{n}=\mu_n$, $f_{10}=g_V = 1$ and $f_{20}=g_M = \mu_p - \mu_n-1$, $N_i$,  $\bar{N}_i$,  $n_{ij}$ and $\bar{n}_{ij}$ are some positive integer numbers, and
\begin{align}
G(l^2,\Lambda^{2},n)\equiv\frac{(-\Lambda^2)^{n}}{(l^2-\Lambda^2)^{n}}.
\end{align}
Also we have the constrained conditions as
\begin{align}
\sum_{j=1}^{\bar{N}_i}  b_{ij}&=1, \notag\\
\sum_{j=1}^{N_i}  a_{ij}&=1 ~~~~\text{for}~~~~ i=1,2,4, \notag\\
\sum_{j=1}^{N_i}  a_{3j}&=0.
\label{equation:general-FFs-this-work}
\end{align}

It is easy to verify that these forms of FFs revert to some standard ones with specific values of $N_{i}, \bar{N}_{i}, \bar{n}_{ij}$ and $n_{ij}$. For example, by setting $\bar{N}_4=1, \bar{n}_{41}=2, \bar{\Lambda}_{41}=\Lambda_W$, and $b_{41}=1$, $f_4$ returns to the usual dipole form, as shown in Eq. (\ref{equation:dipole-FFs-of-Wnp}).


\section{Analytic results for the $\gamma W$-exchange contribution}
To discuss the $\gamma W$-exchange contributions in a general form, we perform the calculation at the amplitude level rather than at the cross-section level, as is usually done in the literature. To achieve this aim, we reduce the four-dimensional (4D) amplitudes to a two-dimensional (2D) form and express them in the following manner:
\begin{align}
\mathcal{M}^{X} &\equiv \mathcal{N}\sum_{i=1}^{16} \mathcal{C}_i^{X} O^i,
\end{align}
where
\begin{align}
\mathcal{N} \equiv -\frac{m_N\sqrt{E_\nu(E_e+m_e)}g^2V_{ud}}{4m_W^2},
\end{align}
and $X$ refers to $W$ or $\gamma W$, $\mathcal{C}_i$ are coefficients and $O_i$ are chosen as follows:
\begin{align}
O_1  &\equiv \big[\xi_p^{\dagger} \xi_n \big] \big[ \xi_e^{\dagger} \bm{\sigma}\cdot\bm{n}_e \eta_{\nu}\big],
&O_2 &\equiv \big[\xi_p^{\dagger} \xi_n \big] \big[ \xi_e^{\dagger} \bm{\sigma}\cdot\bm{n}_{\nu}\eta_{\nu}\big], \notag\\
O_3  &\equiv \big[\xi_p^{\dagger} \xi_n \big] \big[ \xi_e^{\dagger} \eta_{\nu}\big],~~~~~~~~~~~~~~~~~~~~~~~~~~
&O_4 &\equiv \big[\xi_p^{\dagger} \xi_n \big] \big[ \xi_e^{\dagger} \bm{\sigma}\cdot(\bm{n}_e\times\bm{n}_{\nu})\eta_{\nu}\big], \notag\\
O_5  &\equiv \big[\xi_p^{\dagger} \bm{\sigma}\cdot\bm{n}_e \xi_n \big] \big[ \xi_e^{\dagger} \eta_{\nu}\big],
&O_6 &\equiv \big[\xi_p^{\dagger} \bm{\sigma}\cdot\bm{n}_{\nu} \xi_n \big] \big[ \xi_e^{\dagger} \eta_{\nu}\big],\notag\\
O_7  &\equiv \big[\xi_p^{\dagger} \bm{\sigma}\cdot(\bm{n}_e\times\bm{n}_{\nu}) \xi_n \big] \big[ \xi_e^{\dagger} \bm{\sigma}\cdot\bm{n}_e \eta_{\nu}\big],
&O_8 &\equiv \big[\xi_p^{\dagger} \bm{\sigma}\cdot(\bm{n}_e\times\bm{n}_{\nu}) \xi_n \big] \big[ \xi_e^{\dagger} \bm{\sigma}\cdot\bm{n}_{\nu} \eta_{\nu}\big], \notag\\
O_9  &\equiv  \big[\xi_p^{\dagger} \bm{\sigma}\xi_n \big]\times \big[ \xi_e^{\dagger} \bm{\sigma} \eta_{\nu}\big] \cdot\bm{n}_e,
&O_{10} &\equiv \big[\xi_p^{\dagger} \bm{\sigma}\xi_n \big]\times \big[ \xi_e^{\dagger} \bm{\sigma} \eta_{\nu}\big] \cdot\bm{n}_{\nu},
\notag\\
O_{11}  &\equiv \big[\xi_p^{\dagger} \bm{\sigma} \cdot \xi_n\big]\cdot \big[ \xi_e^{\dagger} \bm{\sigma} \eta_{\nu}\big],
&O_{12} &\equiv \big[\xi_p^{\dagger} \bm{\sigma}\cdot\bm{n}_e \xi_n \big] \big[ \xi_e^{\dagger} \bm{\sigma}\cdot\bm{n}_e \eta_{\nu}\big],
\notag\\
O_{13}  &\equiv \big[\xi_p^{\dagger} \bm{\sigma}\cdot\bm{n}_e \xi_n \big] \big[ \xi_e^{\dagger} \bm{\sigma}\cdot\bm{n}_{\nu} \eta_{\nu}\big],
&O_{14} &\equiv \big[\xi_p^{\dagger} \bm{\sigma}\cdot\bm{n}_{\nu} \xi_n \big] \big[ \xi_e^{\dagger} \bm{\sigma}\cdot\bm{n}_e \eta_{\nu}\big]
\notag\\
O_{15}  &\equiv \big[\xi_p^{\dagger} \bm{\sigma}\cdot\bm{n}_{\nu} \xi_n \big] \big[ \xi_e^{\dagger} \bm{\sigma}\cdot\bm{n}_{\nu} \eta_{\nu}\big],
&O_{16} &\equiv \big[\xi_p^{\dagger} \bm{\sigma}\cdot(\bm{n}_e\times\bm{n}_{\nu}) \xi_n \big] \big[ \xi_e^{\dagger}  \eta_{\nu}\big].
\end{align}
Here $\xi,\eta$ are the Pauli spinors of particle and anti-particle with corresponding helicity, and $\bm{\sigma}$ is the Pauli matrix. The unit vectors $\bm{n}_e$ and $\bm{n}_{\nu}$ are oriented along the direction of the electron's three-momentum $\bm{p}_e$ and antineutrino's three-momentum $\bm{p}_{\nu}$, respectively.

Furthermore, one can separate the amplitudes into  Fermi (F) part and  Gamow-Teller (GT) part as
\begin{align}
\mathcal{M}^{X} &\equiv  \mathcal{M}_{\text{F}}^{X}+ \mathcal{M}_{\text{GT}}^{X},\notag\\
\mathcal{M}_{\text{F}}^{X} &\equiv \mathcal{N}\sum_{i=1}^4 \mathcal{C}_i^{X} O^i, \notag\\
\mathcal{M}_{\text{GT}}^{X} & \equiv \mathcal{N}\sum_{i=5}^{16} \mathcal{C}_i^{X} O^i.
\end{align}

In the practical calculation, we approximately take the antineutrino mass as $m_{\nu} \approx 0$ and use the package FeynCalc~\cite{FeynCalc}   to handle the Dirac matrices in four dimensions, the package PackageX~\cite{PackageX-v3.0} to perform the loop integration, and the package LoopTools~\cite{LoopTools} for double-checking.

For the one-$W$-exchange contribution, we  first calculate the amplitude beyond the FAL  and then expand the coefficients $\mathcal{C}_{i}^{W}$ in terms of $m_W^{-1}$ and $m_n^{-1}$ to the leading order (LO). Finally we get
\begin{gather}
\mathcal{C}_{2,\text{LO}}^{W} =-\mathcal{C}_{3,\text{LO}}^{W} = g_V, \notag\\
\mathcal{C}_{6,\text{LO}}^{W} =i\mathcal{C}_{10,\text{LO}}^{W}=-\mathcal{C}_{11,\text{LO}}^{W} =  g_A,
\end{gather}
and other 11 coefficients are zero at LO.

For the $\gamma W$-exchange contribution, we first apply the FAL to the 4D amplitude, excluding the Dirac spinors, before performing the loop integration. Next, we reduce the 4D amplitude to its 2D form, keeping  $\boldsymbol{p}_e, \boldsymbol{p}_{\nu}$ and $m_e$ in Dirac spinors  as non-zero. We then utilize the package PackageX \cite{PackageX-v3.0} to conduct the loop integration. Finally, we treat $m_W^{-1}$, $|\boldsymbol{p}_{e}|$, $ |\boldsymbol{p}_{\nu}|$, and $m_e$ as small quantities and expand $\mathcal{C}_i^{\gamma W}$ in terms of these small variables to LO. Similarly, we find that only $\mathcal{C}_{2,3,6,10,11}^{\gamma W}$ are non-zero at LO in the small variables, and they can be expressed as:
\begin{align}
\mathcal{C}_{2,\text{LO}}^{\gamma W} =&\frac{\alpha_eg_A}{8\pi} [d_{2,1}F_{10}^{p} + d_{2,2}F_{20}^{p}  + d_{2,3} F_{10}^{n} +d_{2,4}  F_{20}^{n}] ,\notag\\
\mathcal{C}_{6,\text{LO}}^{\gamma W} =&\frac{\alpha_eg_V}{8\pi}[d_{6,1}^{V}F_{10}^{p} + d_{6,2}^{V}F_{20}^{p}  + d_{6,3}^{V} F_{10}^{n} +d_{6,4}^{V}  F_{20}^{n}]\notag\\
+&\frac{\alpha_eg_M}{8\pi}[d_{6,1}^{M}F_{10}^{p} + d_{6,2}^{M}F_{20}^{p}+d_{6,3}^{M} F_{10}^{n} +d_{6,4}^{M}  F_{20}^{n}],
\end{align}
where
\begin{align}
d_{2,i}&= \sum_{j,k}\hat{{\cal F}}_{ij,4k}\Big[\frac{X_1(\Lambda_{ij},\Lambda_{4k})}{2m_N^4(\Lambda_{ij}^2-\Lambda_{4k}^2)} - \frac{\Lambda_{ij}Z_1(\Lambda_{4k})-\Lambda_{4k}Z_1(\Lambda_{ij})}{m_N^4\Lambda_{ij}\Lambda_{4k}(\Lambda_{ij}^2-\Lambda_{4k}^2)}\Big],
\label{equation:expressions-for-d2}
\end{align}
and
\begin{align}
d_{6,1}^{V}&= \sum_{j,k}\hat{{\cal F}}_{1j,1k}
\Big[\frac{X_2(\Lambda_{1j},\Lambda_{1k})}{6m_N^4(\Lambda_{1j}^2-\Lambda_{1k}^2)} + \frac{\Lambda_{1j}Z_2(\Lambda_{1k})-\Lambda_{1k}Z_2(\Lambda_{1j})}{3m_N^4\Lambda_{1j}\Lambda_{1k}(\Lambda_{1j}^2-\Lambda_{1k}^2)}\Big],\notag\\
d_{6,2}^{V}&=\sum_{j,k}\hat{{\cal F}}_{2j,1k}
\Big[\frac{X_3(\Lambda_{2j},\Lambda_{1k})}{3m_N^4(\Lambda_{2j}^2-\Lambda_{1k}^2)} + \frac{2[
\Lambda_{2j}Z_3(\Lambda_{1k})-\Lambda_{1k}Z_3(\Lambda_{2j})]}{3m_N^4\Lambda_{2j}\Lambda_{1k}(\Lambda_{2j}^2-\Lambda_{1k}^2)}\Big],\notag\\
d_{6,3}^{V}&=[d_{6,1}^{V}~\textrm{after replacing the index $1j$ with $3j$}],\notag\\
d_{6,4}^{V}&=[d_{6,2}^{V}~\textrm{after replacing the index $2j$ with $4j$}],
\end{align}
and
\begin{align}
d_{6,1}^{M}&=[d_{6,2}^{V}~\textrm{after replacing the indexes $2j$ and $1k$ with $1j$ and $2k$, respectively}],\notag\\
d_{6,2}^{M}&= \sum_{j,k}\hat{{\cal F}}_{2j,2k}
\Big[\frac{X_4(\Lambda_{2j},\Lambda_{2k})}{4m_N^4(\Lambda_{2j}^2-\Lambda_{2k}^2)} - \frac{
Z_4(\Lambda_{2k})-Z_4(\Lambda_{2j})}{m_N^4(\Lambda_{2j}^2-\Lambda_{2k}^2)}\Big],\notag\\
d_{6,3}^{M}&=[d_{6,1}^{M}~\textrm{after replacing the index $1j$ with $3j$}],\notag\\
d_{6,4}^{M}&=[d_{6,2}^{M}~\textrm{after replacing the index $2j$ with $4j$}].
\label{equation:expressions-for-d2}
\end{align}
Here the operator $\hat{{\cal F}}_{ij,mk}$ is defined as
\begin{align}
\hat{{\cal F}}_{ij,mk}\equiv a_{ij}b_{mk} \Lambda_{ij}^{2n_{ij}}\bar{\Lambda}_{mk}^{2\bar{n}_{mk}}\frac{(-1)^{n_{ ij}+\bar{n}_{mk}}}{(n_{ ij}-1)!(\bar{n}_{mk}-1)!}\frac{d^{n_{ ij}-1}}{d(\Lambda_{ij}^2)^{n_{ij}-1}}\frac{d^{\bar{n}_{mk}-1}}{d(\bar{\Lambda}_{mk}^2)^{\bar{n}_{mk}-1}},
\end{align}
and the functions $X_i,Y,Z_i$ are defined as
\begin{align}
X_1(x,y) &\equiv x^2\log\frac{m_N^2}{x^2}-y^2\log\frac{m_N^2}{y^2}+6m_N^2\log\frac{x^2}{y^2},\notag\\
X_2(x,y) &\equiv -x^2\log\frac{m_N^2}{x^2}+y^2\log\frac{m_N^2}{y^2}+6m_N^2\log\frac{x^2}{y^2},\notag\\
X_3(x,y) &\equiv x^2\log\frac{m_N^2}{x^2}-y^2\log\frac{m_N^2}{y^2},\notag\\
X_4(x,y) &\equiv -2x^2\log\frac{m_N^2}{x^2}+2y^2\log\frac{m_N^2}{y^2}-m_N^2\log\frac{x^2}{y^2},\notag\\
Y(x)     &\equiv\log[\frac{x + \sqrt{-4 m_N^2 + x^2}}{2 m_N}],\notag\\
Z_1(x) &\equiv (-4 m_N^2 + x^2)^{3/2}Y(x),\notag\\
Z_2(x) &\equiv (-4 m_N^2 + x^2)^{1/2}(8m_N^2 + x^2)Y(x),\notag\\
Z_3(x) &\equiv (-4 m_N^2 + x^2)^{1/2}(2m_N^2 + x^2)Y(x),\notag\\
Z_4(x) &\equiv (-4 m_N^2 + x^2)^{1/2}xY(x).
\end{align}

In addition, the following relations are also obtained:
\begin{align}
d_{3,j}=-d_{2,j},~~d_{10,j}= -i d_{6,j},~~d_{11,j}= -d_{6,j},
\end{align}
with $j=1,2,3,4$.

We would like to note that for denominators such as $\Lambda_{ij}^2-\Lambda_{4k}^2$ in the expressions $d_{ij}$, the limit $\Lambda_{ij}\rightarrow\Lambda_{4k}$ is understood when $\Lambda_{ij}=\Lambda_{4k}$.

From the above expression, we can define the $\gamma W$-exchange corrections as:
\begin{align}
 \frac{\alpha_e}{2\pi}\delta_i &\equiv \frac{{\cal C}_{i,\text{LO}}^{\gamma W}}{{\cal C}_{i,\text{LO}}^W}.
\end{align}
It is easy to verify that the corrections $C_{\text{Born}}^{\text{F/GT}}$
defined in Refs. \cite{Towner-1992,Hayen-2021} can be expressed as
\begin{align}
C_{\text{Born}}^{\text{F}}=\delta_2 &= \delta_3 , \notag\\
C_{\text{Born}}^{\text{GT}}=\delta_6 &= \delta_{10}=\delta_{11}.
\end{align}
For simplicity, we also separate $C_{\text{Born}}^{\text{F,GT}}$ into two parts as:
\begin{align}
C_{\text{Born}}^{\text{F}} &\equiv  C_{\text{Born}}^{\text{F},g_A} + C_{\text{Born}}^{\text{F},g_M},\notag\\
C_{\text{Born}}^{\text{GT}} &\equiv  C_{\text{Born}}^{\text{GT},g_V} + C_{\text{Born}}^{\text{GT},g_M},
\end{align}
where the indexes $g_{A,V,M}$ refer to the contributions from $g_{A,V,M}$ in the $\gamma W$-exchange contributions, respectively. We would like to mention that $C_{\text{Born}}^{\text{F},g_M}$ is zero and we will not be considered this in the following discussion.

\section{Numerical comparison and discussions}
\subsection{Numerical inputs}
To  provide a clear picture of the $\gamma W$-exchange corrections, we present the numerical results in this section where the following masses and coupling constants are used
\begin{align}
&m_n = 939.56542~\text{MeV}, m_p=938.27209~\text{MeV}, m_e=0.51100~\text{MeV},\notag\\
&F_{10}^{p}=1,F_{20}^{p}=1.793,F_{20}^{n}=-1.913,\notag\\
&g_V=1, g_A=-1.26, g_M =F_{20}^{p}-F_{20}^{n}=3.706.
\end{align}

For the EM FFs of proton and neutron, $F_{1,2}^{p,n}$, perturbative theory suggests that
\begin{align}
&F_{1}^{p,n} (Q^2\rightarrow \infty) \propto  \frac{1}{Q^4},\notag\\
&F_{2}^{p,n} (Q^2\rightarrow \infty) \propto  \frac{1}{Q^6},
\label{equation:pQCD-behavior1}
\end{align}
so we choose the following $N_{i}$ and corresponding $n_{ij}$ in Eq.~(\ref{equation:general-FFs-this-work}) as
\begin{align}
&N_{i}=2,~n_{11,12,31,32}=2,~n_{21,22,41,42}=3.
\label{equation:EM-FFs-form-1}
\end{align}
We fit the parameters $a_{ij}$ and $\Lambda_{ij}$ in $F_{1,2}^{p}$ using the experimental datasets from Ref.~\cite{Arrington:2007ux} and fit the corresponding parameters in $F_{1,2}^{n}$ using the formulas from Ref.~\cite{ZhiHongYe-2018}. Finally, we choose the parameters as follows:
\begin{align}
&N_{1}=2, n_{1j}=2, a_{11}=0.609,\Lambda_{11}=0.815,a_{12}=0.391,\Lambda_{12}=1.230,\notag\\
&N_{2}=2, n_{2j}=3, a_{21}=0.858,\Lambda_{21}=0.970,a_{22}=0.142,\Lambda_{22}=1.598,\notag\\
&N_{3}=2, n_{3j}=2, a_{31}=1, \Lambda_{31}=1.288, a_{32}=-1, \Lambda_{32}=1.378,F_{10}^{n}=1,\notag\\
&N_{4}=2, n_{4j}=3, a_{41}=0.348, \Lambda_{41}=0.699, a_{42}=0.652, \Lambda_{42}=1.214,
\label{equation:EM-FFs-parameters-1}
\end{align}
where $j=1,2$, the unit of $\Lambda_{ij}$ is GeV while $a_{ij}$ is dimensionless.

We would like to mention that the $\gamma W$-exchange contributions from the elastic intermediate state are not sensitive to the  behavior of FFs in very high energy, therefore we do not apply the following constraint condition:
\begin{align}
&\mu_pG_{E}^{p}/G_M^{p}  \xrightarrow{Q^2\rightarrow \infty}1,
\label{equation:pQCD-behavior1}
\end{align}
where
\begin{align}
G_M^{p,n}(Q^2)&\equiv F_1^{p,n}(Q^2)+F_2^{p,n}(Q^2),\notag\\
G_E^{p,n}(Q^2)&\equiv F_1^{p,n}(Q^2)-\frac{Q^2}{4m_{p,n}^2} F_2^{p,n}(Q^2).
\end{align}

For comparison, we also use the following simple forms  for the EM FFs as inputs to present the results
\begin{align}
&N_{1}=1, n_{11}=2, a_{11}=1,\Lambda_{11}=1.006,\notag\\
&N_{2}=1, n_{21}=3, a_{21}=1,\Lambda_{21}=1.123,\notag\\
&N_{3}=2, n_{3j}=1, a_{31}=1, \Lambda_{31}=0.847, a_{32}=-1, \Lambda_{32}=0.914, F_{10}^{n}=1,\notag\\
&N_{4}=1, n_{41}=3, a_{41}=1, \Lambda_{41}=1.038.
\label{equation:EM-FFs-parameters-2}
\end{align}

Naively, the EM FFs used in Ref.~\cite{Towner-1992} correspond to the following choices:
\begin{align}
&N_{i}=1, n_{i1}=2, a_{i1}=1,\Lambda_{i1}=\Lambda_{\gamma}=0.84, F_{10}^{n}=0,
\label{equation:EM-FFs-parameters-3}
\end{align}
with $i=1,2,3,4$.

\begin{figure*}[htb]
\centering
\includegraphics[height=6cm]{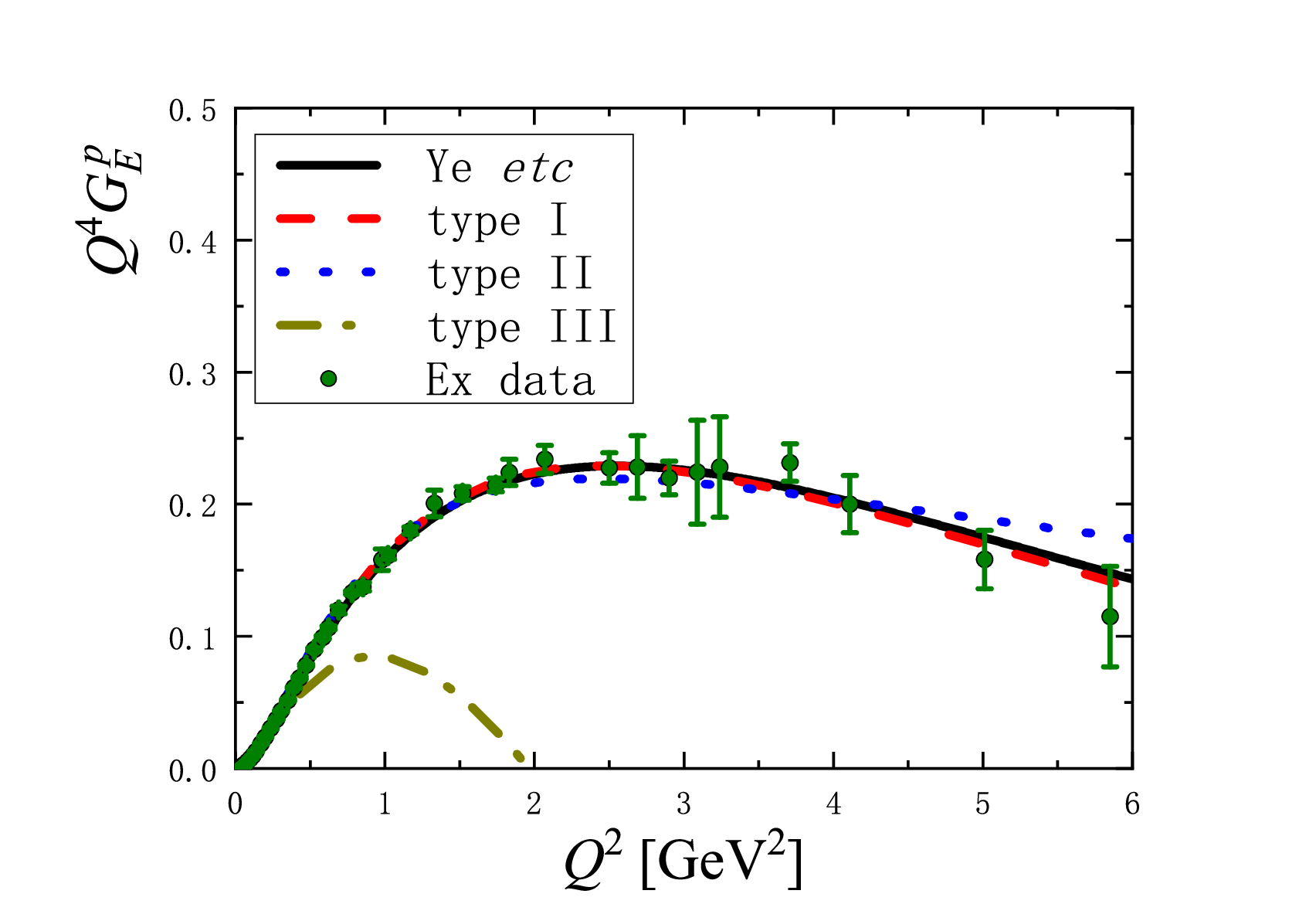}\includegraphics[height=6cm]{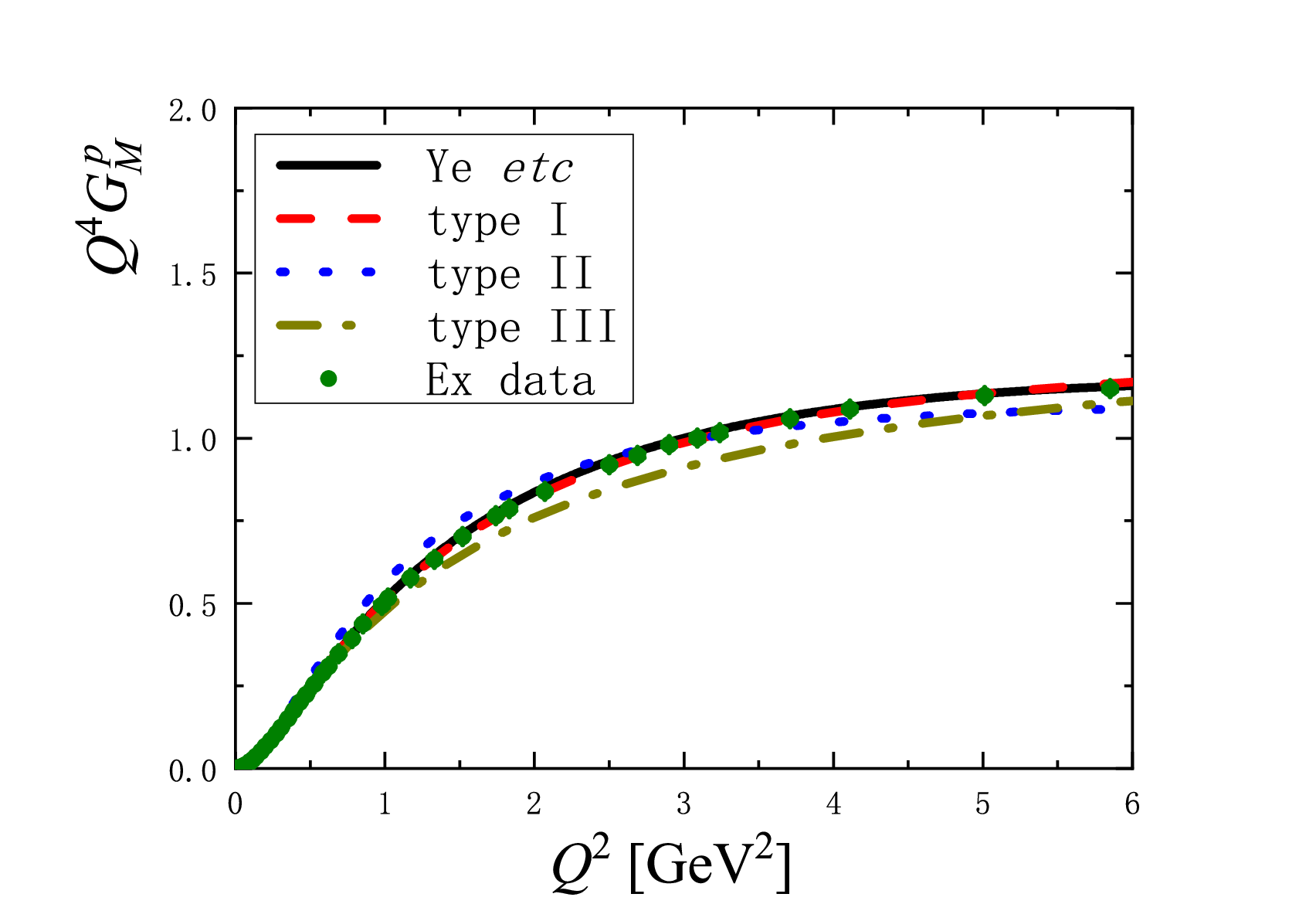}
\includegraphics[height=6cm]{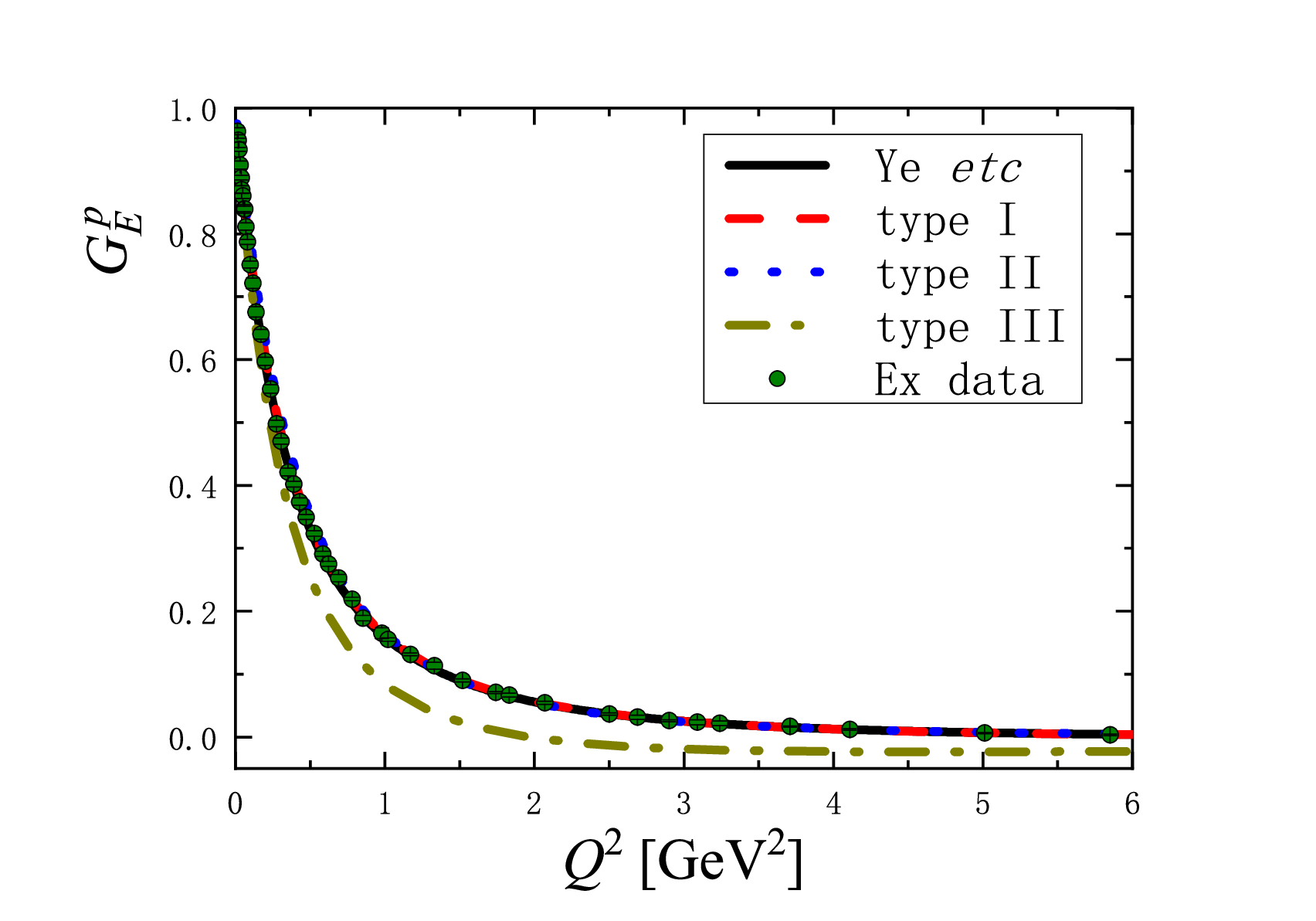}\includegraphics[height=6cm]{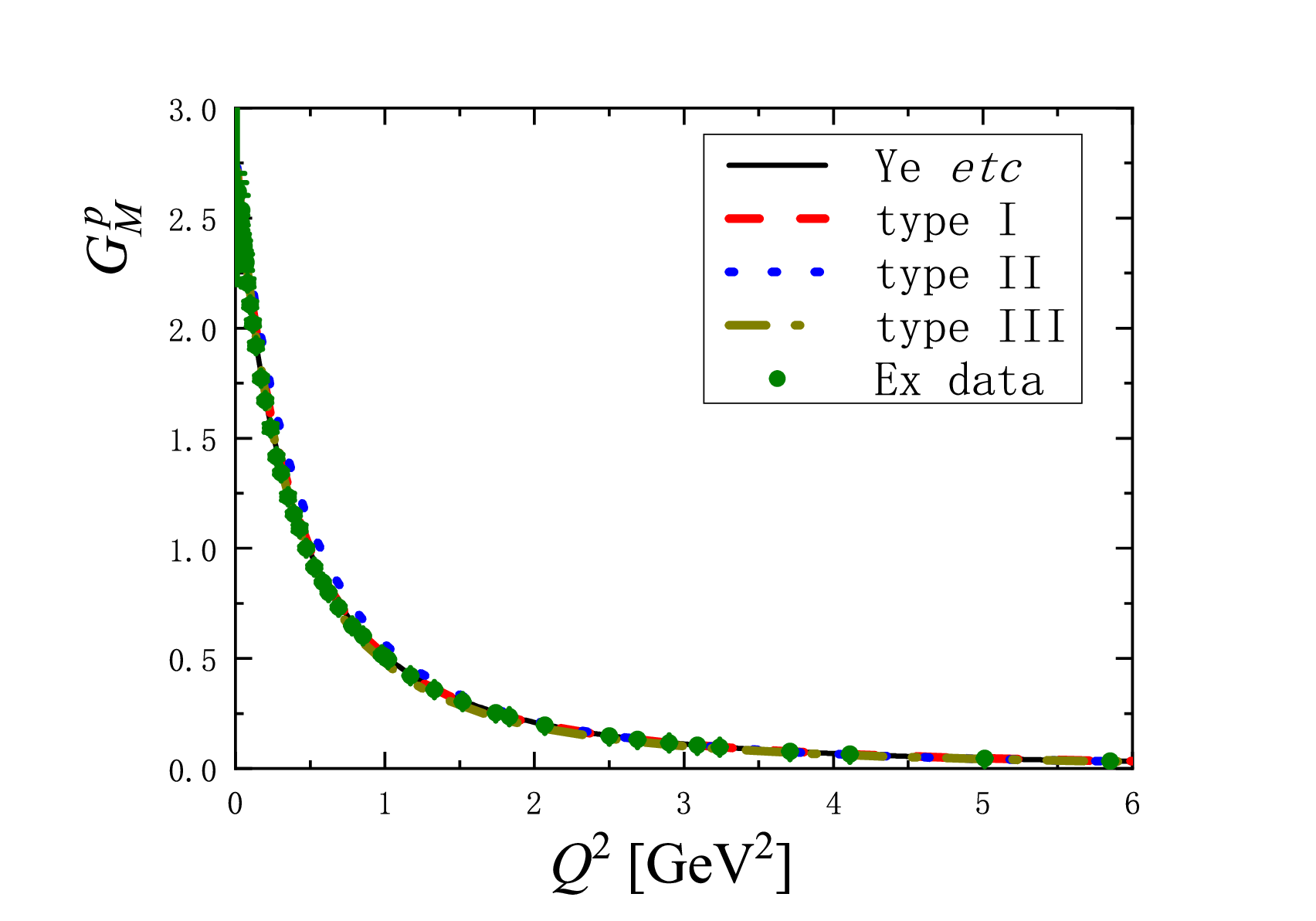}
\caption{Numerical comparison for the EM FFs of proton $G_{E,M}^{p}$ with different parameters. The above panel is for $Q^4G_{E,M}^{p}(Q^2)$ vs. $Q^2$ and the bottom panel is for $G_{E,M}^{p}(Q^2)$ vs. $Q^2$. The results labelled by Ye {\it etc} refer to the fitted results in Ref.~\cite{ZhiHongYe-2018}. The type $\RM1,\RM2,\RM3$ are corresponding to the choices of parameters by Eqs. (\ref{equation:EM-FFs-parameters-1}, \ref{equation:EM-FFs-parameters-2}, \ref{equation:EM-FFs-parameters-3}), respectively.  The experimental data sets for $G_{E,M}^{p}$ are taken from Ref.~\cite{Arrington:2007ux}.
}
\label{figure:GE-and-GM-1}
\end{figure*}

\begin{figure*}[htb]
\centering
\includegraphics[height=6cm]{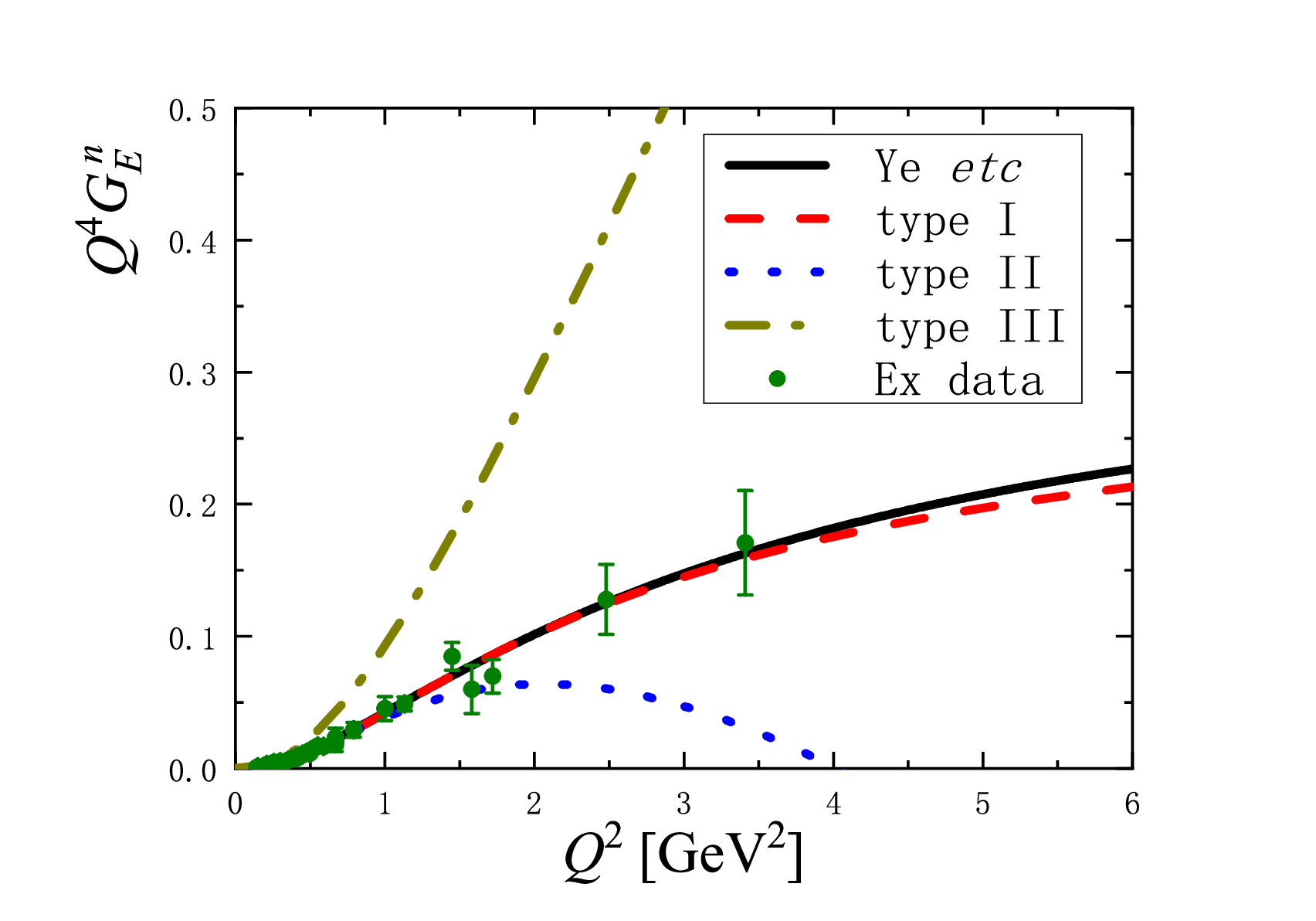}\includegraphics[height=6cm]{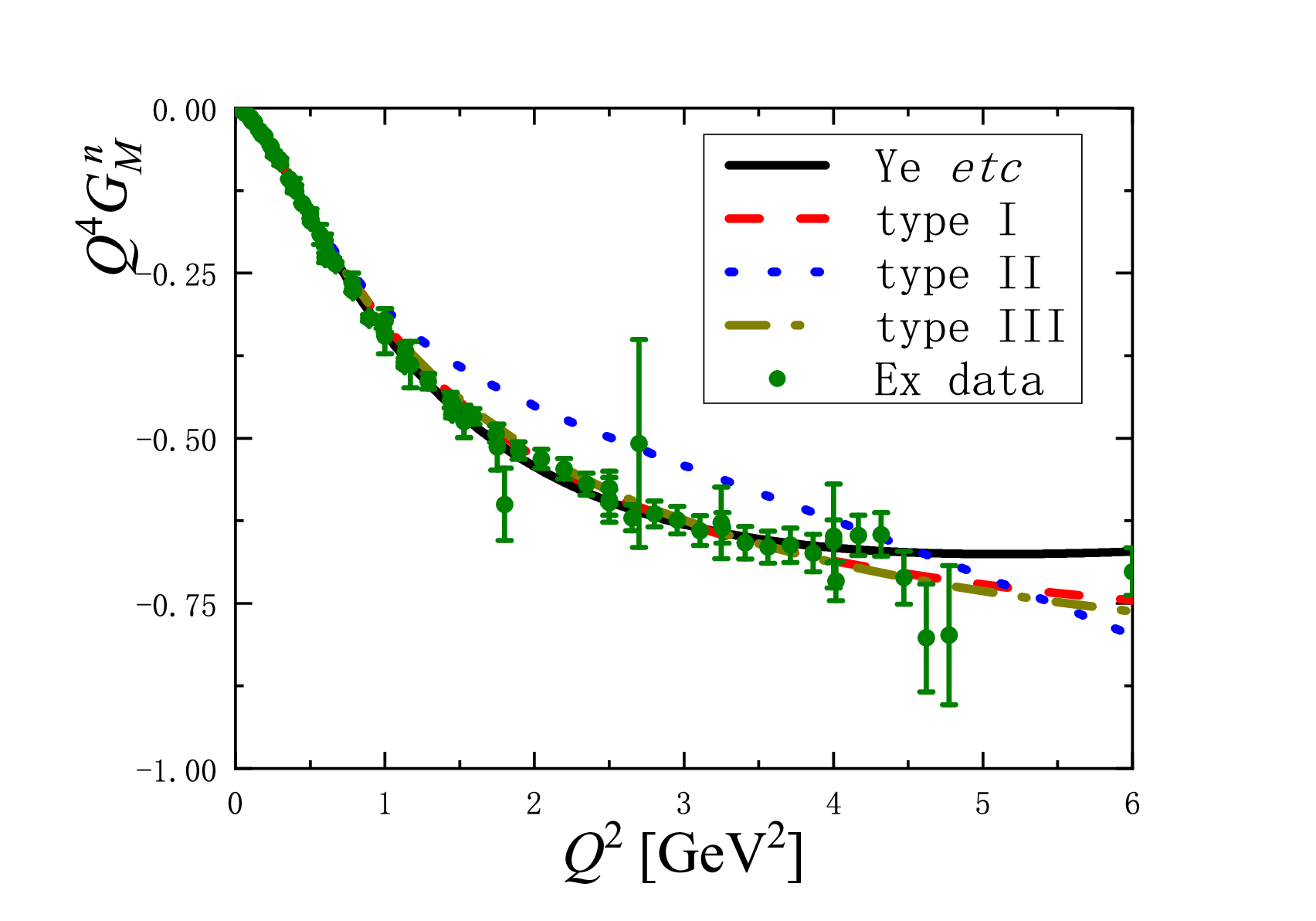}
\includegraphics[height=6cm]{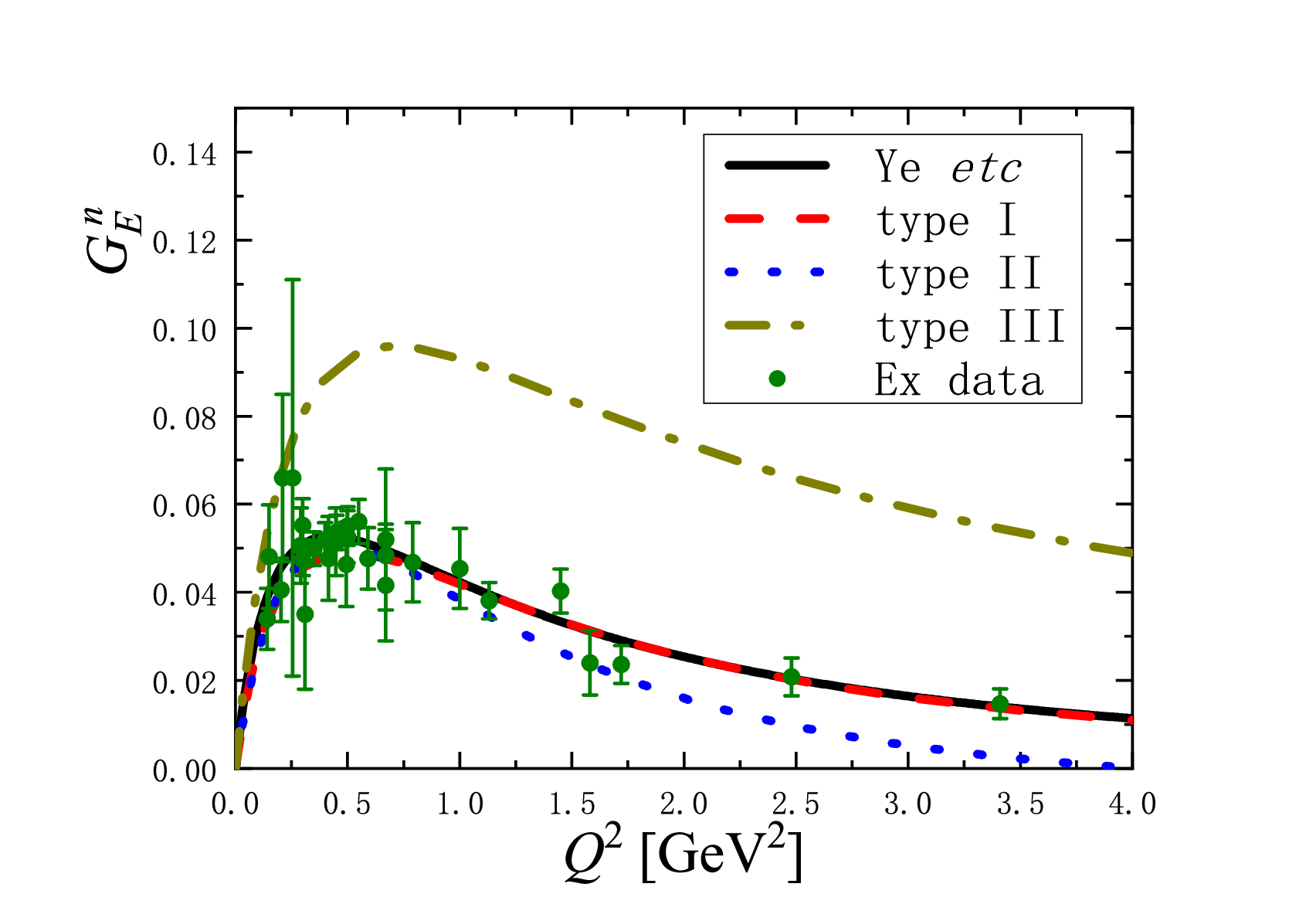}\includegraphics[height=6cm]{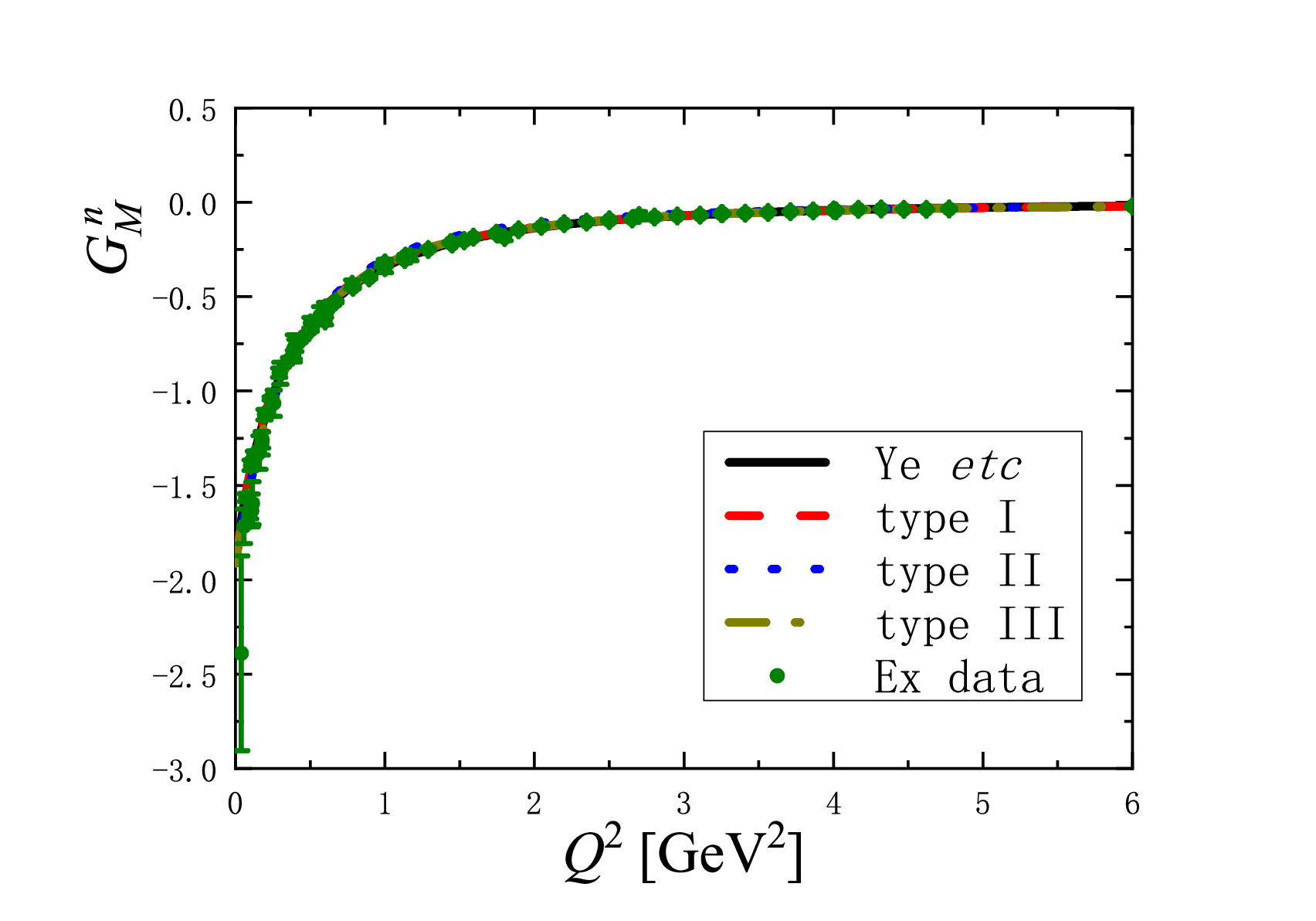}
\caption{Numerical comparison for the EM FFs of neutron $G_{E,M}^{n}$ with different parameters. The above panel is for $Q^4G_{E,M}^{n}(Q^2)$ vs. $Q^2$ and the bottom panel is for $G_{E,M}^{n}(Q^2)$ vs. $Q^2$. The results labelled by Ye {\it etc} refer to the fitted results in Ref.~\cite{ZhiHongYe-2018}. The type $\RM1,\RM2,\RM3$ are corresponding to the choices of parameters by Eqs. (\ref{equation:EM-FFs-parameters-1}, \ref{equation:EM-FFs-parameters-2}, \ref{equation:EM-FFs-parameters-3}), respectively. The  data sets for $G_{E}^{n}(Q^2)$ are taken from Refs. ~\cite{GE-neutron-Ex-1,GE-neutron-Ex-2,GE-neutron-Ex-3,GE-neutron-Ex-4,GE-neutron-Ex-5,GE-neutron-Ex-6,GE-neutron-Ex-7,GE-neutron-Ex-8,
GE-neutron-Ex-9,GE-neutron-Ex-10,GE-neutron-Ex-11,GE-neutron-Ex-12,GE-neutron-Ex-13,GE-neutron-Ex-14,GE-neutron-Ex-15}. The experimental data sets for $G_{M}^{n}(Q^2)$ are taken from Refs. ~\cite{GM-neutron-Ex-1,GM-neutron-Ex-2,GM-neutron-Ex-3,GM-neutron-Ex-4,GM-neutron-Ex-5, GM-neutron-Ex-6,GM-neutron-Ex-7,GM-neutron-Ex-8,
GM-neutron-Ex-9,GM-neutron-Ex-10,GM-neutron-Ex-11,GM-neutron-Ex-12,GM-neutron-Ex-13,GM-neutron-Ex-14}.
}
\label{figure:GE-and-GM-2}
\end{figure*}

\begin{figure*}[htb]
\centering
\includegraphics[height=6cm]{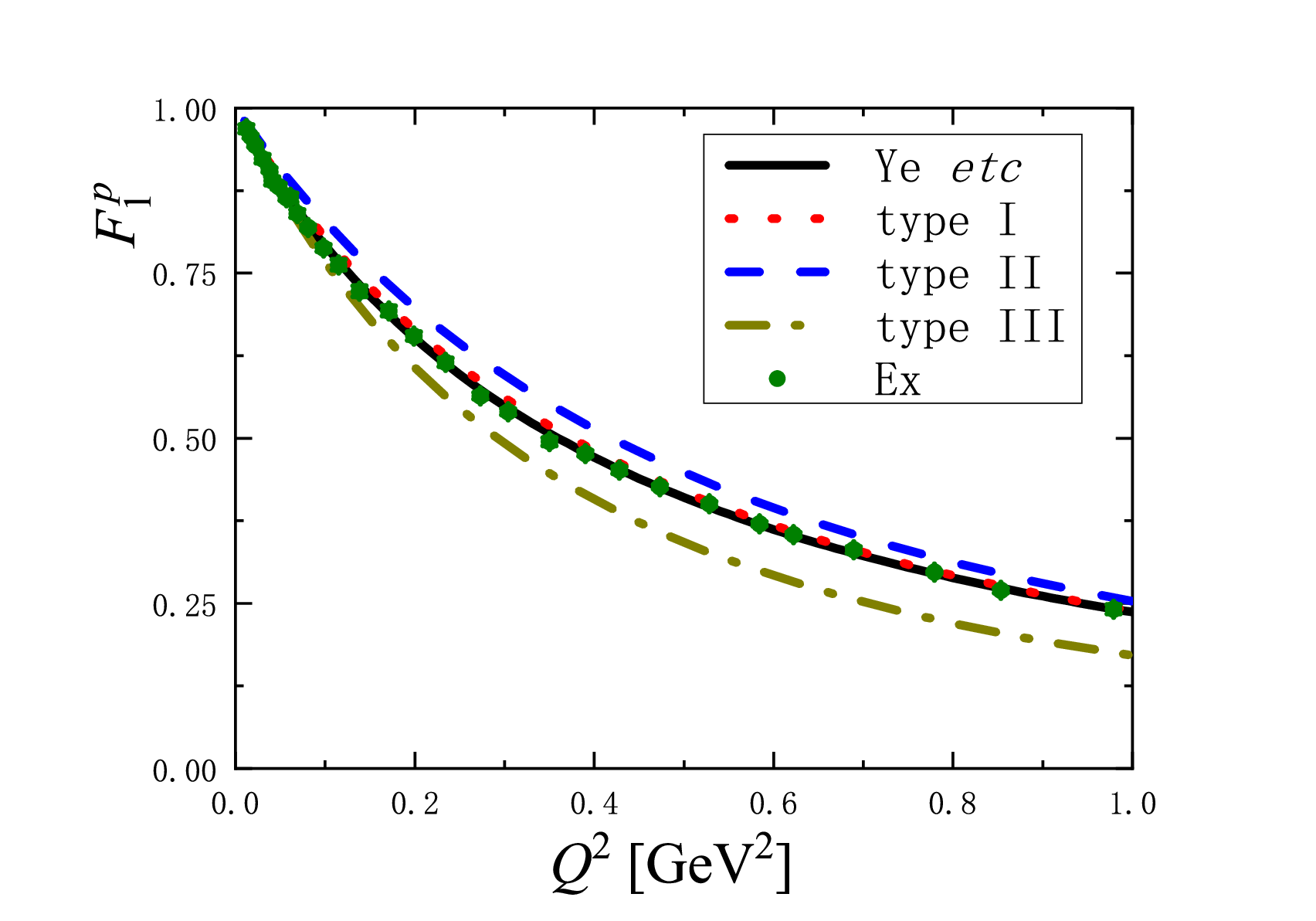}\includegraphics[height=6cm]{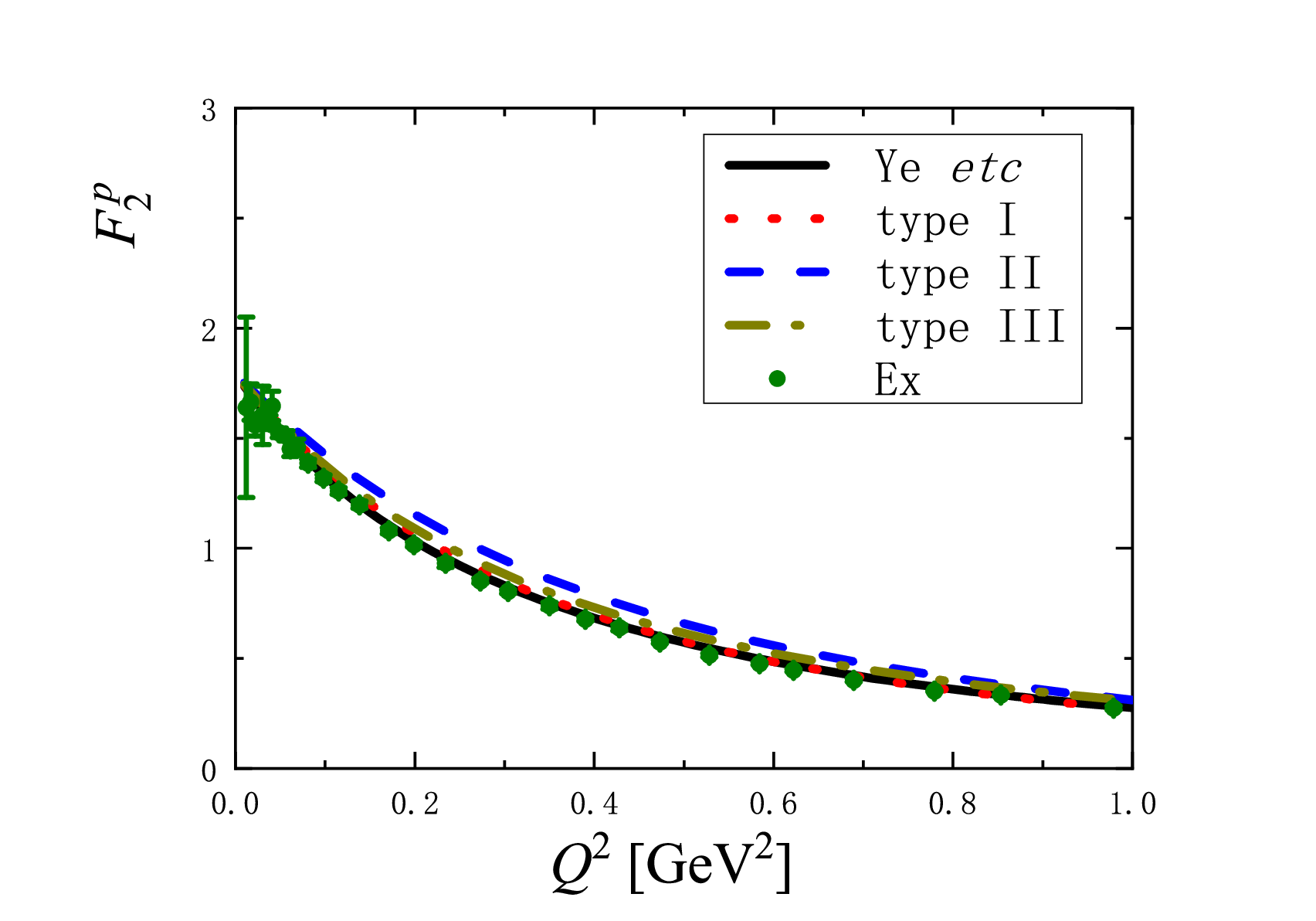}
\includegraphics[height=6cm]{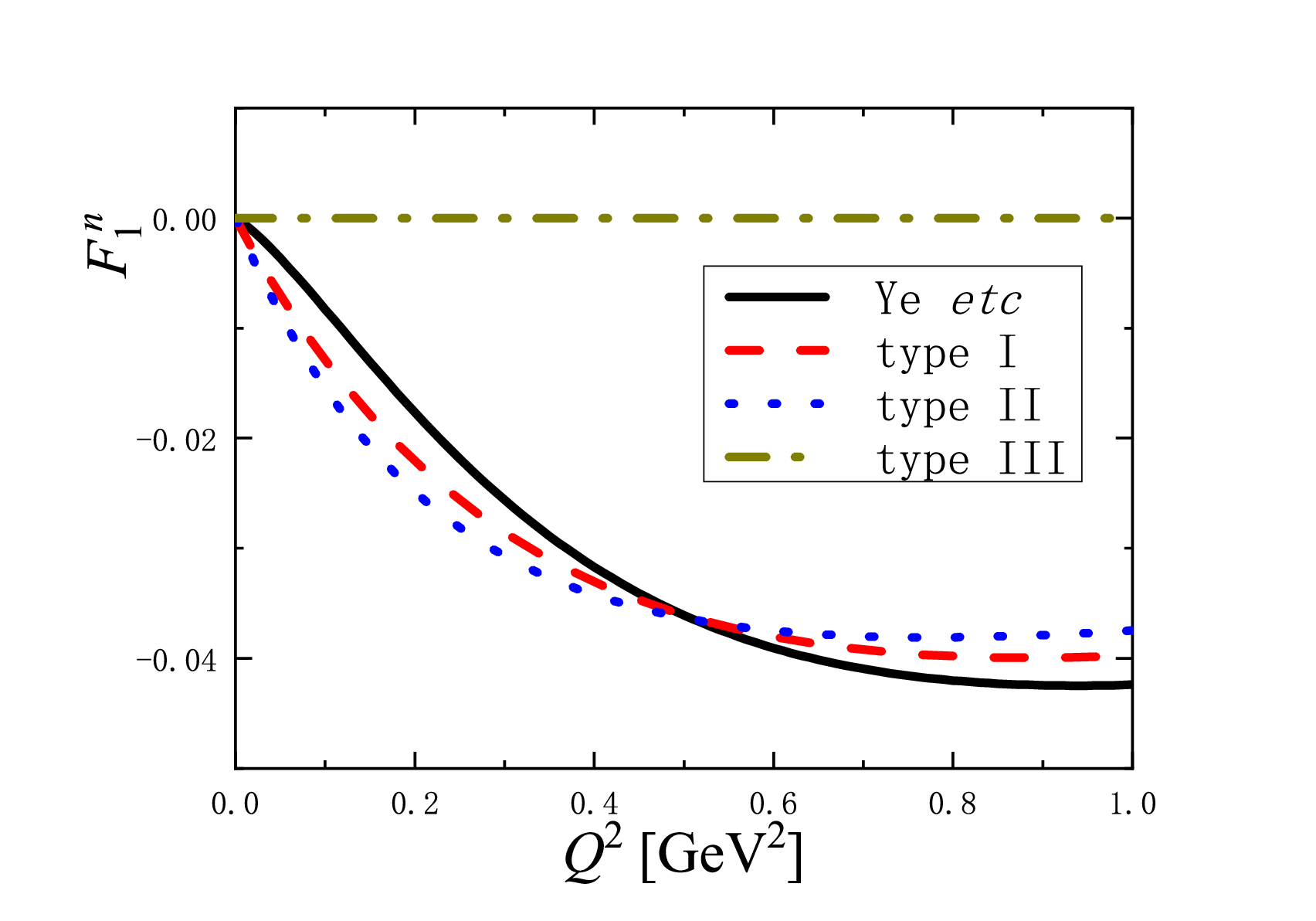}\includegraphics[height=6cm]{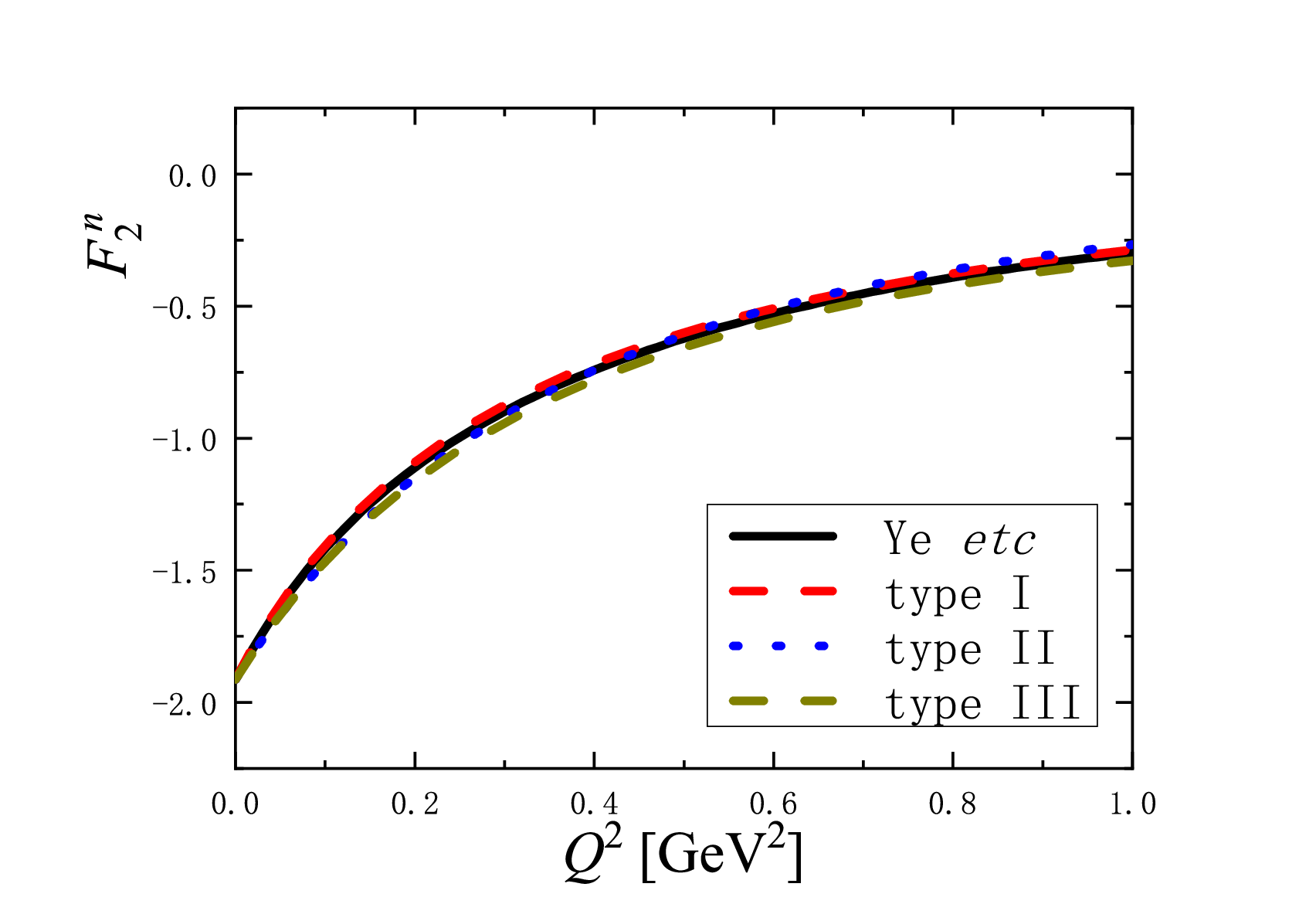}
\caption{Numerical comparison for the EM FF $F_{1,2}^{p,n}$ with different parameters. The above panel is for $F_{1,2}^{p}(Q^2)$ vs. $Q^2$ and the bottom panel is for $F_{1,2}^{n}$ vs. $Q^2$. The results labelled by Ye {\it etc} refer to the fitted results in Ref.~\cite{ZhiHongYe-2018}. The type $\RM1,\RM2,\RM3$ are corresponding to the choices of parameters by Eqs. (\ref{equation:EM-FFs-parameters-1}, \ref{equation:EM-FFs-parameters-2}, \ref{equation:EM-FFs-parameters-3}), respectively.  The data sets for $F_{1,2}^{p}$ are combined from the experimental data sets $G_{E,M}^{p}(Q^2)$ from Ref.~\cite{Arrington:2007ux}. The data sets for $F_{1,2}^{n}$ are not presented.
}
\label{figure:F1F2}
\end{figure*}

In the following, we name the choices of the EM FFs and cut-offs by Eqs.~(\ref{equation:EM-FFs-parameters-1},~\ref{equation:EM-FFs-parameters-2},~\ref{equation:EM-FFs-parameters-3}) as types $\RM1,\RM2$ and $\RM3$, respectively. A detailed comparison of the corresponding EM FFs $G_{E,M}^{p,n}$ and the fits in  Ref.~\cite{ZhiHongYe-2018} is presented in Figs.~\ref{figure:GE-and-GM-1},~\ref{figure:GE-and-GM-2}, where the world datasets for  $G_{E,M}^{p}$ \cite{Arrington:2007ux}, $G_E^n$ \cite{GE-neutron-Ex-1,GE-neutron-Ex-2,GE-neutron-Ex-3,GE-neutron-Ex-4,GE-neutron-Ex-5,GE-neutron-Ex-6,GE-neutron-Ex-7,GE-neutron-Ex-8,
GE-neutron-Ex-9,GE-neutron-Ex-10,GE-neutron-Ex-11,GE-neutron-Ex-12,GE-neutron-Ex-13,GE-neutron-Ex-14,GE-neutron-Ex-15} and $G_M^n$ \cite{GM-neutron-Ex-1,GM-neutron-Ex-2,GM-neutron-Ex-3,GM-neutron-Ex-4,GM-neutron-Ex-5,GM-neutron-Ex-6,GM-neutron-Ex-7,GM-neutron-Ex-8,
GM-neutron-Ex-9,GM-neutron-Ex-10,GM-neutron-Ex-11,GM-neutron-Ex-12,GM-neutron-Ex-13,GM-neutron-Ex-14} are also included.
The comparisons in Fig.~\ref{figure:GE-and-GM-1} show that, globally, the $G_{E, M}^{p}$ in types $\RM1$ and $\RM2$ and the fits in  Ref.~\cite{ZhiHongYe-2018} are highly consistent with experimental datasets in the region $Q^2\subset[0, 6] \text{ GeV}^2$, while the $G_{E,M}^{p}$ in type $\RM3$ is slightly underestimated. The comparisons in Fig.~\ref{figure:GE-and-GM-2} indicate that the $G_{E,M}^{n}$ in type $\RM1$ and the fits in  Ref.~\cite{ZhiHongYe-2018} are consistent with the experimental datasets, while the $G_{E,M}^{n}$ in types $\RM2$ and $\RM3$ show some discrepancies.  Since, in practical calculations, the EM FFs $F_{1,2}^{p,n}$ are used and the results are somewhat sensitive to their low energy-behaviors, we show the corresponding behaviors in Fig.~\ref{figure:F1F2}. In this figure we can see that the FFs $F_{1,2}^{p}$ and $F_{2}^{n}$ in type~{\RM1} are almost the same as the fits in  Ref.~\cite{ZhiHongYe-2018}, while the other results differ slightly from those fits in  Ref.~\cite{ZhiHongYe-2018}. We will demonstrate that these differences lead to variations in the corresponding $C_{\text{Born}}$.

For the weak FFs $f_{1,2}(Q^2)$, we assume the isospin symmetry and use Eq.~(\ref{equation:f12-by-EM-FFs}) to derive $f_{1,2}(Q^2)$ from the EM FFs $F_{1,2}^{p,n}(Q^2)$. Once $F_{1,2}^{p,n}(Q^2)$ are known, all the parameters in $f_{1,2}(Q^2)$, such as $b_{ij}$ and $\bar{\Lambda}_{ij}$, are fixed. The only unknown is $f_4(Q^2)$. Since the experimental datasets for $f_{4}(Q^2)$ are insufficient, we simply take them to be the same as those used in Ref.~\cite{Towner-1992}, which corresponds to the following special choices in Eq.~(\ref{equation:general-FFs-this-work}):
\begin{align}
\bar{N}_4=1, \bar{n}_{41}=2, b_{41}=1, \bar{\Lambda}_{41}=\Lambda_{W}=1.09\pm0.05~\text{GeV}.
\label{equation:weak-FFs-parameters}
\end{align}
In principle, one can choose more general forms for $f_4$ using Eq.~(\ref{equation:general-FFs-this-work}) when the experimental datasets are sufficiently robust. Here, we simply take it as an example for comparison.

\subsection{Numerical Results for $C_{\text{Born}}$}
By using the types $\RM1,\RM2$ and $\RM3$ FFs as inputs, we list the final numerical results for $C_{\text{Born}}$ in Tab. \ref{table:CBorn-num}, where we have maintained a precision of $0.1\%$. However, we would like to emphasize that the recoil contributions beyond the FAL in the $\gamma W$-exchange will result in a difference of approximately parts-per-thousand. In Ref.~\cite{Towner-1992}, the parameters used in FFs  are the same as those in Eq.~(\ref{equation:EM-FFs-parameters-3}). In Ref.~\cite{Hayen-2021}, the global fit results of Ref. \cite{ZhiHongYe-2018} for nucleon magnetic Sachs FFs and the model-independent z-expansion analysis of Ref.~\cite{Bhattacharya-gA} for the axial form factor were utilized.  For comparison, we presented these two related results and our results in Tab.~\ref{table:CBorn-num}.

\begin{table}[htbp]
\renewcommand\arraystretch{1}
\centering
\begin{tabular}{p{2cm}<{\centering}p{4cm}<{\centering}p{3cm}<{\centering}p{3cm}<{\centering}p{3cm}<{\centering}}
\hline
\hline
&FFs & $C_{\text{Born}}^{\text{F},g_A}$ & $C_{\text{Born}}^{\text{GT},g_V}$ & $C_{\text{Born}}^{\text{GT},g_M}$ \\
\hline
Ref.~\cite{Towner-1992}
& Eqs. (\ref{equation:general-FFs-this-work}, \ref{equation:weak-FFs-parameters}, \ref{equation:EM-FFs-parameters-2})
& $0.881\pm 0.014$
&no calculated
&no calculated\\
\hline
\hline
Ref.~\cite{Hayen-2021}
&Refs. \cite{ZhiHongYe-2018,Bhattacharya-gA}
& 0.91(5)
& 0.39(1)
&0.78(2)  \\
\hline
\hline
type $\RM1$
&Eqs. (\ref{equation:general-FFs-this-work}, \ref{equation:weak-FFs-parameters}, \ref{equation:EM-FFs-parameters-1})
&$0.951$
&$0.442$
&$0.835$\\

\hline
type $\RM2$
&Eqs. (\ref{equation:general-FFs-this-work}, \ref{equation:weak-FFs-parameters}, \ref{equation:EM-FFs-parameters-2})
&$1.009$
&$0.475$
&$0.876$\\
\hline
type $\RM3$
&Eqs. (\ref{equation:general-FFs-this-work}, \ref{equation:weak-FFs-parameters}, \ref{equation:EM-FFs-parameters-3})
& $0.882$
&$0.388$
&$0.832$\\
\hline\hline
\end{tabular}
\caption{The numerical results for $C_{\text{Born}}$ in Refs. \cite{Towner-1992,Hayen-2021} and this work.
}
\label{table:CBorn-num}
\end{table}

In Tab.~\ref{table:CBorn-num}, one can see that the results for $C_{\text{Born}}^{\text{F},g_A}$ with types $\RM1$ and $\RM3$ EM FFs are consistent with those given in Refs.~\cite{Towner-1992,Hayen-2021} within the error, while the corresponding result with type $\RM2$ EM FFs are slightly larger.   In contrast to the case of $C_{\text{Born}}^{\text{F},g_A}$, our results (type $\RM1$) for $C_{\text{Born}}^{\text{GT},g_V}$ and $C_{\text{Born}}^{\text{GT},g_M}$ are about 13\% and 7\% larger than those reported in Ref.~\cite{Hayen-2021}, respectively. The comparisons of the results for $C_{\text{Born}}^{\text{GT},g_V}$ from types $\RM1$, $\RM2$, and $\RM3$ suggest that the results are somewhat sensitive to the choices of the EM FFs. The comparisons of the results for $C_{\text{Born}}^{\text{GT},g_M}$ from types $\RM1,\RM2,\RM3$ with that in Ref. \cite{Hayen-2021} indicate that their difference is not due to the choices of FFs. A possible reason for this discrepancy is the use of the FCC approximation.  As discussed above, the naive FCC approximation is non-unique. In our practical calculations, we also choose different forms of the FCC approximation as examples for comparison and find that the results vary.

To present the contributions more clearly, we also separate the contributions from different parts, which are expressed as follows:
 \begin{equation}
\text{type \RM1}:
\left\{
\begin{aligned}
C_{\textrm{Born}}^{\textrm{F},g_A}&=1.058F_{10}^{p}+0.985F_{20}^{p}-0.027F_{10}^n+0.966F_{20}^{n} =0.951, \\
C_{\textrm{Born}}^{\textrm{GT},g_V}&=0.473F_{10}^{p}+0.238F_{20}^{p}-0.013F_{10}^{n}+0.233F_{20}^{n}=0.442, \\
C_{\textrm{Born}}^{\textrm{GT},g_M}&=0.853F_{10}^{p}-0.019F_{20}^{p}-0.019F_{10}^{n}-0.019F_{20}^{n}=0.835,
\end{aligned}
\right.
\label{equation:results-by-EM-FFs-parameters-1}
\end{equation}
and
\begin{equation}
\text{type \RM2}:
\left\{
\begin{aligned}
C_{\textrm{Born}}^{\textrm{F},g_A}&=1.084F_{10}^{p}+1.024F_{20}^{p}-0.028F_{10}^n+0.985F_{20}^{n} =1.009, \\
C_{\textrm{Born}}^{\textrm{GT},g_V}&=0.498F_{10}^{p}+0.255F_{20}^{p}-0.014F_{10}^{n}+0.244F_{20}^{n}=0.475, \\
C_{\textrm{Born}}^{\textrm{GT},g_M}&=0.901F_{10}^{p}-0.024F_{20}^{p}-0.021F_{10}^{n}-0.020F_{20}^{n}=0.876,
\end{aligned}
\right.
\label{equation:results-by-EM-FFs-parameters-2}
\end{equation}
and
\begin{equation}
\text{type \RM3}:
\left\{
\begin{aligned}
C_{\textrm{Born}}^{\textrm{F},g_A}&=0.999F_{10}^{p}+0.999F_{20}^{p}+0.999F_{20}^{n}=0.882, \\
C_{\textrm{Born}}^{\textrm{GT},g_V}&=0.414F_{10}^{p}+0.224F_{20}^{p}+0.224F_{20}^{n}=0.388, \\
C_{\textrm{Born}}^{\textrm{GT},g_M}&=0.829F_{10}^{p}-0.024F_{20}^{p}-0.024F_{20}^{n}=0.832.
\end{aligned}
\right.
\label{equation:results-by-EM-FFs-parameters-3}
\end{equation}

The results in Eqs.~(\ref{equation:results-by-EM-FFs-parameters-1},~\ref{equation:results-by-EM-FFs-parameters-2},~\ref{equation:results-by-EM-FFs-parameters-3}) show  that the EM FF $F_{1}^{p}$ contributes the most significantly to $C_{\text{Born}}^{\text{GT}}$. The comparison among these results indicates that they are somewhat sensitive to the cutoff in  $F_{1}^{p}$. The differences can be traced to the varying behaviors of $F_{1,2}^{p,n}$ presented in Fig.~\ref{figure:F1F2}. For example, the $F_{1}^{p}$ in type $\RM2$ case is  slightly larger than that in type $\RM1$, which results in the contributions from type  $\RM2$ being larger than those from $\RM1$ and $\RM3$. There is about $5\%-13\%$ difference due to the different parameterizations of the EM FFs. In this work, we favor the type $\RM1$ case.

Furthermore, we list the relative ratios of the contributions from different EM FFs in Tab. \ref{table:CBorn-ratio}. It clearly shows that  contributions from $F_{1}^{p}, F_{2}^{p,n}$ are significant  for $C_{\text{Born}}^{\text{F},g_A}$ and $C_{\text{Born}}^{\text{GT},g_V}$. Additionally, there are substantial cancellations between $F_{2}^{p}$ and $F_{2}^{n}$. The contributions from $F_{1}^{n}$ are of the same order as the sum of $F_{2}^{p}$ and $F_{2}^{n}$, accounting for a few percent of the full corrections. These properties  indicate that the precise measurements of $F_{1,2}^{n}$ are crucial for achieving an accurate estimation of the $\gamma W$-exchange contributions. Another interesting observation is that the contributions to $C_{\text{Born}}^{\text{GT},g_M}$ primarily come from $F_{1}^{p}$.

\begin{table}[htbp]
\renewcommand\arraystretch{1}
\centering
\begin{tabular}{p{2cm}<{\centering}p{4cm}<{\centering}p{3cm}<{\centering}p{3cm}<{\centering}p{3cm}<{\centering}}
\hline
\hline
contributions/All& $C_{\text{Born}}^{\text{F},g_A}(f_4)$ & $C_{\text{Born}}^{\text{GT},g_V}(f_1)$ & $C_{\text{Born}}^{\text{GT},g_M}(f_2)$ \\
\hline
$F_{1}^{p}$
& $111\%$
& $107\%$
&$102\%$\\
\hline
\hline
$F_{2}^{p}$
& $186\%$
& $97\%$
& $-4\%$   \\
\hline
\hline
$F_{1}^{n}$
& $-3\%$
& $-3\%$
& $-2\%$  \\
\hline
\hline
$F_{2}^{n}$
& $-194\%$
& $-100\%$
& $4\%$   \\
\hline
\hline
\end{tabular}
\caption{The ratioes of the different contributions in type $\RM1$ case.}
\label{table:CBorn-ratio}
\end{table}

\begin{figure*}[htb]
\centering
\includegraphics[height=8cm]{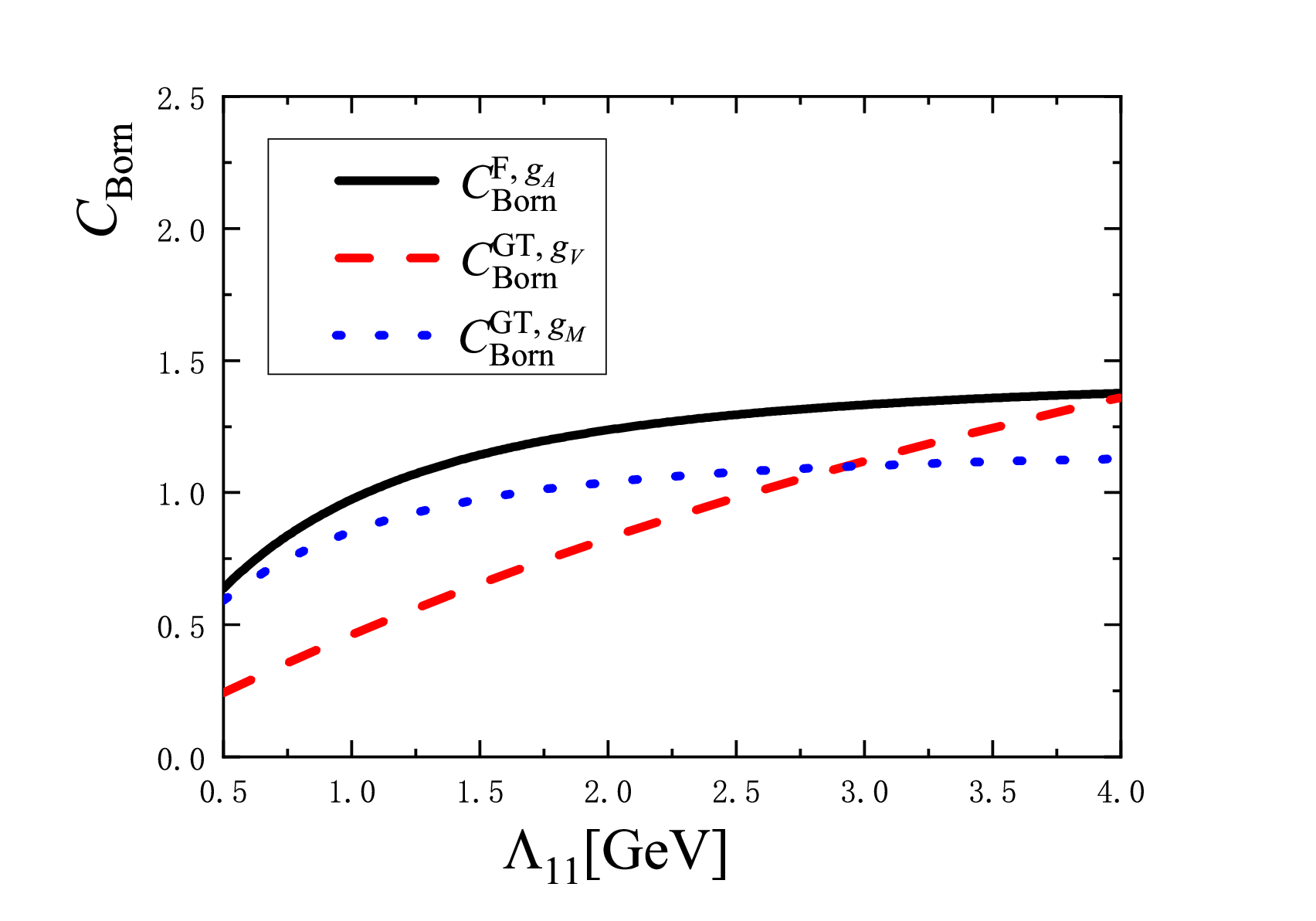}
\caption{Numeric results for $C_{\text{Born}}$ vs. $\Lambda_{11}$ where the type $\RM1$ EM FFs and $\Lambda_{12}=\Lambda_{11}$ are used.}
\label{figure:delta-Lambda1}
\end{figure*}

In Fig. \ref{figure:delta-Lambda1}, we present the numerical results for the dependence of $C_{\text{Born}}^{\text{F,GT}}$ on $\Lambda_{11}$ in type $\RM1$ case with $\Lambda_{12}=\Lambda_{11}$.  It can be observed that the results are somewhat sensitive to the cutoff. Naively, when the cutoff $\Lambda_{12}=\Lambda_{11}\rightarrow \infty$, the results with $F_{20}^{p}=0$ should revert to the point-like particle case. The large difference between the results with the physical $\Lambda_{ij}$ and $\infty$ indicates that, although neutron $\beta$ decay is a low energy process, the internal structure remains crucial for estimating the $\gamma W$-exchange contribution. This property also suggests that the naive power counting of the low-energy effective theory should be applied with caution in neutron $\beta$ decay.

\section{Summary}

In summary, the $\gamma W$-exchange contributions with the elastic intermediate state in neutron $\beta$ decay are discussed at the amplitude level in the FAL. The non-uniqueness in the usual FAL and FCC approximations is addressed. To avoid this non-uniqueness, we first perform the full calculation at the amplitude level and then separate the contributions into inner and outer parts.

The analytical results for the $\gamma W$-exchange contribution are presented with a general form for the EM FFs of the proton and neutron as inputs. The parameters in this general form of the form factors can be determined by future precise experimental data sets, and the derived analytical expressions are helpful for analyzing the uncertainty.

By fitting with the current experimental datasets for the EM FFs, our numeric results for $C_{\text{Born}}^{\text{F},g_A}$ are consistent with those presented in Refs.~\cite{Towner-1992,Hayen-2021}. Additionally, our numerical results for $C_{\text{Born}}^{\text{GT},gV}$ and $C_{\text{Born}}^{\text{GT},g_M}$ are approximately 13\% and 7\% larger than those reported in Ref.~\cite{Hayen-2021}.  The contributions from different parts of the EM FFs are also discussed, and these results indicate that to achieve a precise estimation of the $\gamma W$-exchange contribution, the precise EM FFs for neutron is needed.

\section{Acknowledgments}

H.~Q.~Zhou would like to thank Zhi-Hui Guo for helpful discussions. H.~Q.~Zhou is supported by the National Natural Science Foundation of China under Grants Nos.~12150013 and 12075058. H.~Y.~Cao is supported by the Science and Technology Research Project of Hubei Provincial Education Department under Grants No.~Q20222502 and by the National Natural Science Foundation of China under Grants No.~12405153.

\bibliography{Refs-TBE-in-neutron-beta-decay-FAL}

\end{document}